\documentclass[fleqn]{article}
\usepackage[LGR,T1]{fontenc}
\usepackage[margin=1.2in]{geometry}
\usepackage{mathtools}
\usepackage{amssymb}
\usepackage{mathrsfs}
\usepackage{dsfont}
\usepackage{amsthm}
\usepackage{newtxtext}
\usepackage{newtxmath}
\usepackage{microtype}
\usepackage{booktabs}
\usepackage{enumitem}
\usepackage[colorlinks=true,linkcolor=blue!50!black,citecolor=blue!50!black,urlcolor=blue!50!black]{hyperref}
\usepackage[most]{tcolorbox}
\usepackage{tikz}
\usepackage[edges]{forest}
\usetikzlibrary{positioning,arrows.meta}
\tcbuselibrary{listings}
\lstdefinelanguage{Lean}{
 morekeywords={import,namespace,end,open,section,variable,omit,in,abbrev,def,lemma,theorem,
  structure,where,instance,inductive,fun,by,have,show,from,let,calc,match,with,if,then,else,
  Type,Prop,noncomputable,set,obtain,rintro,intro,exact,refine,apply,rw,simp,simpa,ext,
  constructor,cases,induction,omega,ring,decide,norm_num,congr,change,funext,unfold,
  fin_cases,split_ifs,by_cases,first,all_goals,done,subst,rcases,use,tauto,group},
 sensitive=true,
 morecomment=[l]{--},
 morestring=[b]",
 morekeywords={absurd,at,attribute,by_contra,exfalso,left,only,push_neg,revert,rfl,right,rwa,universe,using},
 morecomment=[n]{/-}{-/},
}
\newtcbinputlisting{\leanlisting}[2]{
 listing file={#1},listing only,listing options={language=Lean},breakable,enhanced,
 colback=gray!4,colframe=gray!50,boxrule=0.4pt,arc=1pt,
 left=4pt,right=4pt,top=2pt,bottom=2pt,
 title={\ttfamily #2},fonttitle=\small,coltitle=black,colbacktitle=gray!15,
}
\newtcbinputlisting{\leanexcerpt}[4]{
	listing file={#1},listing only,
	listing options={language=Lean,firstline=#2,lastline=#3},
	breakable,enhanced,
	colback=gray!4,colframe=gray!50,boxrule=0.4pt,arc=1pt,
	left=4pt,right=4pt,top=2pt,bottom=2pt,
	title={\ttfamily #4\hfill\normalfont lines~#2--#3},fonttitle=\small,coltitle=black,colbacktitle=gray!15,
}

\usepackage{acro}
\usepackage[noabbrev]{cleveref}
\crefname{equation}{equation}{equations}
\Crefname{equation}{Equation}{Equations}
\usepackage[table]{xcolor}
\usepackage[textsize=tiny,backgroundcolor=white]{todonotes}
\usepackage[normalem]{ulem}

\makeatletter
\g@addto@macro\bfseries{\boldmath}
\makeatother
\DeclareSymbolFont{upgreek}{LGR}{cmr}{m}{n}
\DeclareMathSymbol{\deltaup}{\mathord}{upgreek}{`d}
\DeclareMathSymbol{\piup}{\mathord}{upgreek}{`p}
\DeclareMathSymbol{\epsilonup}{\mathord}{upgreek}{`e}
\DeclareMathSymbol{\omegaup}{\mathord}{upgreek}{`w}
\DeclareMathSymbol{\phiup}{\mathord}{upgreek}{`f}
\definecolor{c1}{RGB}{0,119,187}
\definecolor{c2}{RGB}{51,187,238}
\definecolor{c3}{RGB}{0,153,136}
\definecolor{c4}{RGB}{238,119,51}
\definecolor{c5}{RGB}{204,51,17}
\definecolor{c6}{RGB}{238,51,119}
\definecolor{c0}{RGB}{187,187,187}

\allowdisplaybreaks

\newcommand{\ZNR}[1]{\ensuremath{\mathds{Z}_{#1}^R}}

\newcommand{\Z}[1]{\ensuremath{\mathds{Z}_{#1}}}
\newcommand{\SU}[1]{\ensuremath{\mathrm{SU}(#1)}}

\newcommand{\U}[1]{\ensuremath{\mathrm{U}(#1)}}
\newcommand{\GSM}{\ensuremath{G_{\mathrm{SM}}}}

\newcommand{\ModulusU}{\ensuremath{\tau}}
\newcommand{\ModularWeight}{\ensuremath{k}}
\newcommand{\Generator}[1]{\ensuremath{\mathsf{#1}}}
\newcommand{\Group}[2]{\ensuremath{\text{#1}(#2)}}
\newcommand{\rep}[1]{\boldsymbol{#1}}
\newcommand{\crep}[1]{\overline{\boldsymbol{#1}}}
\newcommand{\lean}[1]{\texttt{#1}}
\newcommand{\SL}{\ensuremath{\Group{SL}{2,\mathds{Z}}}}
\newcommand{\Mod}[1]{\mathcal{M}_{#1}}
\newcommand{\ee}{\mathrm{e}}
\newcommand{\ii}{\operatorname{i}}
\let\defeq\coloneq\let\eqdef\eqcolon
\DeclareMathOperator{\im}{Im}

\DeclareMathOperator{\tr}{tr}
\DeclareMathOperator{\ad}{ad}
\DeclareMathOperator{\Weyl}{Weyl}

\theoremstyle{plain}
\newtheorem{theorem}{Theorem}

\newtheorem{corollary}{Corollary}
\crefname{lemma}{lemma}{lemmas}
\Crefname{lemma}{Lemma}{Lemmas}
\crefname{corollary}{corollary}{corollaries}
\Crefname{corollary}{Corollary}{Corollaries}
\theoremstyle{remark}

\crefname{remark}{remark}{remarks}
\Crefname{remark}{Remark}{Remarks}
\newlist{assumptions}{enumerate}{1}
\setlist[assumptions]{label=(A\arabic*),ref=(A\arabic*),leftmargin=*}
\crefname{assumptionsi}{assumption}{assumptions}
\Crefname{assumptionsi}{Assumption}{Assumptions}

\tcbset{
 statement/.style={enhanced,breakable,sharp corners,boxrule=0pt,frame hidden,
  colback=black!4,borderline west={1pt}{0pt}{black!65},
  left=7pt,right=7pt,top=3pt,bottom=3pt,before skip=\medskipamount,after skip=\medskipamount},
 proofbox/.style={enhanced,breakable,blanker,borderline west={0.4pt}{0pt}{black!35},
  left=7pt,top=0pt,bottom=0pt,before skip=\medskipamount,after skip=\medskipamount},
}
\tcbset{nosep/.style={before upper={\topsep=0pt\partopsep=0pt}}}
\tcolorboxenvironment{theorem}{statement,nosep}
\tcolorboxenvironment{proposition}{statement,nosep}
\tcolorboxenvironment{lemma}{statement,nosep}
\tcolorboxenvironment{corollary}{statement,nosep}
\tcolorboxenvironment{proof}{proofbox,nosep}

\DeclareAcronym{4D}{
  short = 4D,
  long = four-dimensional,
  pdfcomment = four-dimensional
}

\DeclareAcronym{AI}{
  short = AI,
  long = artificial intelligence,
  pdfcomment = artificial intelligence
}

\DeclareAcronym{FN}{
  short = FN,
  long = Froggatt--Nielsen \cite{Froggatt:1978nt},
  pdfcomment = Froggatt--Nielsen
}

\DeclareAcronym{GUT}{
  short = GUT,
  long = grand unified theory,
  long-plural-form = grand unified theories,
  pdfcomment = grand unified theory
}

\DeclareAcronym{MSSM}{
  short = MSSM,
  long = minimal supersymmetric standard model,
  pdfcomment = minimal supersymmetric standard model
}

\DeclareAcronym{QFT}{
  short = QFT,
  long = quantum field theory,
  pdfcomment = quantum field theory
}

\DeclareAcronym{SM}{
  short = SM,
  long = standard model,
  pdfcomment = standard model
}

\DeclareAcronym{SUGRA}{
  short = SUGRA,
  long = supergravity,
  pdfcomment = supergravity
}

\DeclareAcronym{SUSY}{
  short = SUSY,
  long = supersymmetry,
  pdfcomment = supersymmetry
}

\DeclareAcronym{VEV}{
  short = VEV,
  long = vacuum expectation value,
  pdfcomment = vacuum expectation value
}

\DeclareAcronym{VVMF}{
  short = VVMF,
  long = vector-valued modular form,
  pdfcomment = vector-valued modular form
}

\DeclareMathOperator{\rank}{rank}
\DeclareMathOperator{\ind}{ind}
\DeclareMathOperator{\coker}{coker}

\AddToHook{cmd/appendix/before}{\crefalias{section}{appendix}}
\begin{document}

\newcommand{\mytitle}{Lean-Verified No-Go Theorems for $R$-Symmetries and Modular Symmetries in 4D GUTs}
\title{\mytitle}

\begin{titlepage}
\vspace*{1.0cm}

\begin{flushright}
UCI--TR--2026--22%
\\
\end{flushright}

\vspace*{2cm}

\begin{center}
\begin{minipage}{0.75\textwidth}
\centering\Large\sffamily\bfseries\linespread{1.3}\selectfont\mytitle\par
\end{minipage}

\vspace{1cm}

\renewcommand*{\thefootnote}{\fnsymbol{footnote}}

\textbf{%
Mu-Chun Chen\rlap{,}$^{1,\,}$\footnote{muchunc@uci.edu}\quad 
Xiang-Gan Liu\rlap{,}$^{1,\,2,\,}$\footnote{xianggal@uci.edu}\quad
Michael Ratz$^{1,\,}$\footnote{mratz@uci.edu}
~and~
Alexander Stewart$^{1,\,}$\footnote{ajstewa1@uci.edu}
}
\\[8mm]
\textit{\small$^1$~Department of Physics and Astronomy, University of California, Irvine, CA 92697-4575 USA}\\
\textit{\small $^2$ Instituto de F\'isica, Universidad Nacional Aut\'onoma de M\'exico, Cd.~de M\'exico, C.P.~04510, M\'exico}
\end{center}

\vspace*{1cm}

\begin{abstract}
    $R$-symmetries are instrumental in supersymmetric model building. 
    However, a no-go theorem restricts us from using them in a large class of four-dimensional (4D) grand unified theories (GUTs). 
    Specifically, it has been shown that a supersymmetric 4D GUT based on SU(5), with finitely many chiral multiplets, cannot be broken spontaneously to the exact minimal supersymmetric standard model (MSSM) while preserving a discrete $Z_M^R$, with $M \geq 3$, or continuous U(1) $R$-symmetry. 
    We use Lean~4 to formalize this proof. 
    In the process, we identify and close a gap in the general case of the published argument. 
    Further, we discuss the implications for GUT models with modular symmetries. 
    We point out that modular symmetries are inevitably $R$-symmetries in a supersymmetric framework that also describes gravity. 
    At certain critical points in moduli space, discrete $R$-symmetries are linearly realized. 
    In the context of 4D GUTs, the no-go theorem then implies that exotic states become massless at these points, provided the GUT-breaking vacuum preserves these symmetries. 
    In the vicinity of these critical points, these exotics are parametrically light, which may spoil gauge coupling unification. 
\end{abstract}
\vspace*{1cm}
\end{titlepage}
\renewcommand*{\thefootnote}{\arabic{footnote}}
\setcounter{footnote}{0}
\acresetall

\section{Introduction}

Lean has revolutionized mathematics \cite{Castelvecchi:2021nature,CommelinTopaz:2024bams,Buzzard:2024bams,Avigad:2024bams}. 
This is because Lean allows us to formalize proofs, and check them automatically. 
Modern \ac{AI} agents are fluent in the Lean~4 syntax, and can even develop proofs. 
When applied to physics, arguably this development has not yet fully taken off. 
On the other hand, checking proofs is a very relevant task since theorems in the physics literature have been found to be not entirely correct when their proofs were formalized \cite{Tooby-Smith:2026kzj}. 
These results were obtained using PhysLib \cite{Tooby-Smith:2024vqu,physlib}. 
The main purpose of this work is to use Lean~4 to check, and extend, the proof of a no-go theorem~\cite{Fallbacher:2011xg} in the physics literature.

This theorem concerns $R$-symmetries and \acp{GUT}.
$R$-symmetries are instrumental in supersymmetric model building. 
For instance, they can be argued to be essential to address the $\mu$ problem~\cite{Hall:2002up,Babu:2002tx,Lee:2010gv,Lee:2011dya,Baer:2018avn,Bhattiprolu:2021rrj} in the \ac{MSSM}. 
\ac{4D} \ac{SUSY} \acp{GUT} (see e.g.~\cite{Raby:2017ucc} for a review) are popular not only because they are motivated by the fact that gauge couplings unify in the \ac{MSSM}~\cite{Dimopoulos:1981yj} but also because matter furnishes complete \SU5 representations. 
However, there is an obstruction that limits the use of $R$-symmetries in \ac{4D} \acp{GUT}. 
In~\cite{Fallbacher:2011xg} it was shown that a \ac{4D} \ac{GUT} with a simple gauge group cannot yield the exact \ac{MSSM} together with an unbroken $\ZNR{M\ge3}$ or $\U1_R$ symmetry. 
This means that,if the \ac{GUT} model has an $R$-symmetry, there are certain exotic states the mass of which is tied to the scale of $R$-symmetry breakdown. 

This paper has two purposes. 
The first is to make the theorem fully rigorous by formalizing it in the Lean~4 proof assistant on top of Mathlib~\cite{Mathlib}. 
Even though PhysLib, which was used in~\cite{Tooby-Smith:2026kzj}, has extended group-theory capabilities \cite{Krippendorf:2026pou}, we work directly with Lean~4~\cite{deMoura:2021lean4}.
As we shall see in \Cref{sec:No-Go_Theorem,app:Lean}, this leads to a somewhat shorter proof than the one provided in \cite{Fallbacher:2011xg}. 
It also closes a gap in the general case of the published argument, see \Cref{sec:gap}.

The second purpose is to discuss the implications for modular symmetries, which are $R$-symmetries in the \ac{SUGRA} framework~\cite{Dixon:1989fj,Ibanez:1992uh}. 
At certain critical points in moduli space, discrete $R$-symmetries are linearly realized. 
The no-go theorem then implies that, in the context of \ac{4D} \acp{GUT}, at these points exotic states beyond the \ac{MSSM} become massless.

The outline of the rest of the paper is as follows. 
In \Cref{sec:No-Go_Theorem}, we state the abstract no-go theorem, and we prove it.
In \Cref{sec:SU5}, we apply the no-go theorem to $\ZNR{M}$ symmetries in \SU5 \acp{GUT}, and we compare the formalized proof with the original one.
\Cref{sec:Larger} is devoted to the extension of the results to larger gauge groups.
In \Cref{sec:extra_dim}, we explain how higher-dimensional models of grand unification evade the no-go theorem.
\Cref{sec:modular} discusses modular symmetries, which are $R$-symmetries in \ac{SUGRA}. 
There we identify the modular $R$-symmetries that are linearly realized, and we discuss modular zeros, which can forbid mass terms even if no such symmetry is present.
Finally, \Cref{sec:conclusions} contains our concluding remarks.
Details of how we use Lean are deferred to two appendices. 
In \Cref{sec:lean}, we summarize what has been formalized in Lean, and what has been assumed.
\Cref{app:Lean} contains the central parts of the Lean code.

\section{No-Go Theorem}
\label{sec:No-Go_Theorem}

In this section, we develop a more general version of the no-go theorem as given in \cite{Fallbacher:2011xg}, using some key observations from~\cite{Goodman:1985bw}.
This general proof, verified in Lean~4, is then used to show that a \ac{4D} $\mathcal{N}=1$ \ac{GUT} with gauge group \SU5 cannot have both an unbroken discrete $R$-symmetry, \ZNR{M}, with $M \neq 1,2$ and a massless spectrum consisting only of the \ac{MSSM} field content plus gauge singlets.
We then comment on the extension of the no-go proof to larger \ac{GUT} gauge groups, and on how higher-dimensional \acp{GUT} avoid the no-go theorem.

\subsection{Setting, Notation, and Hypotheses}

Consider a \ac{4D} \ac{GUT} with gauge group $G$, where $G$ is a compact connected Lie group.
Let $G_0 \subseteq G$ be the closed connected subgroup left unbroken by the vacuum.
We will denote the maximal torus of $G_0$ as $T_0$, and we denote the set of roots of $G_0$, i.e.\ the non-zero $T_0$-weights of $\mathfrak{g}_0 \otimes \mathds{C}$, as $\Delta_0 = \Delta(G_0,T_0)$. 
Note in particular that we do not assume $\rank G_0 = \rank G$, which will be crucial in extending our results beyond \SU5 \ac{GUT} gauge groups later. 

Let $\mathcal{S}$ be the group of linear symmetries of the microscopic theory that commute with $G$, which includes $R$-symmetries that act on the superspace coordinates $\theta$.
The vacuum in general is not preserved by all of $G \times \mathcal{S}$.
We define $\Sigma_0 \subseteq G \times \mathcal{S}$ as the subgroup of $G \times \mathcal{S}$ that leaves the vacuum invariant.
The theorem we develop will apply to any unbroken symmetry element $r \in \Sigma_0$, whose action on $\theta$ generically multiplies $\theta$ by a phase $\zeta$.
We assume that $\zeta^2 \neq 1$ since the transformation that sends all spinors and $\theta$ to their negatives is a symmetry of any relativistic \ac{QFT}, and not a nontrivial $R$-symmetry, see also \Cref{ftn:no_order_1_or_2_R-symmetries}.
We further define $B$ as a closed connected subgroup of $G$, with $G_0 \subseteq B \subseteq G$, that commutes with $r$.
Since $B$ and $r$ commute, they will share a simultaneous eigenspace, which will prove useful later.

The centralizer and normalizer of $T_0$ in $B$ are given as
\begin{subequations}
\begin{align}
	C_B(T_0) &= \bigl\{ b \in B \, \big\vert \, b\,t\,b^{-1} = t \; \forall t \in T_0\bigr\}\;,\\
	 N_B(T_0) &= \bigl\{ b \in B \, \big\vert \, b \,T_0\, b^{-1} = T_0\bigr\}\;,
\end{align}
\end{subequations}
respectively. 
In words, the centralizer $C_B(T_0)$ of $T_0$ in $B$ consists of the elements $b \in B$ that commute with all of $T_0$.
The normalizer $N_B(T_0)$ of $T_0$ in $B$ consists of the elements $b \in B$ that map $T_0$ back to itself.
That is, the normalizer permutes the $T_0$-weights, while elements of $C_B(T_0)$ commute with all of $T_0$ and leave every weight fixed.
Two normalizer elements that differ by a centralizer element then permute the weights in the same way.
Therefore, the part of the Weyl group that preserves the torus of $G_0$ is given by
\begin{subequations}
\begin{align}
	\Weyl(B,T_0) &= N_B(T_0)/C_B(T_0)\;, \\
   (w\cdot \chi)(t) &= \chi(n^{-1}\,t\, n)\;,\quad w \in \Weyl(B,T_0)\;.
\end{align}  
\end{subequations}
Here, $n \in N_B(T_0)$ represents $w$ as a coset representative\footnote{$w \in \Weyl(B,T_0) = N_B(T_0)/C_B(T_0)$ is a coset, i.e.\ $w = n\, C_B(T_0)$.}, $\chi$ is a $T_0$-weight, and $t \in T_0$.
This reduces to the ordinary Weyl group of $G$ when $T_0$ is the maximal torus\footnote{Recall that the maximal torus can be written as $T=\bigl\{\exp(\ii\alpha_1\,\Generator{H}_1+\dots+\ii\alpha_\ell \Generator{H}_\ell)\bigm\vert\alpha_j\in\mathds{R}\bigr\}\cong\U{1}^\ell$, where the $\Generator{H}_i$ are the Cartan generators. 
It is the largest Abelian subgroup of $G$ of the form $\U{1}\times\dots\times\U{1}$.\label{ftn:maximal_torus}} of $G$ and $B = G$.
Since $C_B(T_0)$ is connected\rlap{,}\footnote{The centralizer of a torus in a compact connected Lie group is connected.} we can give a more concrete expression for $\Weyl(B,T_0)$ in terms of a maximal torus $T_B$ of $B$ containing $T_0$ as
\begin{equation}\label{eq:weylBT0}
	\Weyl(B,T_0) = \left.\bigl\{ w \in \Weyl(B,T_B) \, \big\vert \, w\,\mathfrak{t}_0 = \mathfrak{t}_0\bigr\}\right\rvert_{\mathfrak{t}_0}\;.
\end{equation}
This just says that in practice we can take the ordinary Weyl group of $B$, keep only the elements that map the Lie algebra $\mathfrak{t}_0$ of $T_0$ to itself, and then restrict them to $\mathfrak{t}_0$.

We now split the massless fermions in the spectrum based on the action of $r$.
We take $r$ to act linearly and unitarily on all fields.
If $r$ acts on the scalar of a chiral multiplet by $\xi$, it acts on the fermion in the same multiplet by $\xi \zeta^{-1}$.
On the gauginos of $B$ sitting in vector multiplets, $r$ acts by $\zeta$.
Define $\mathcal{F}_{\pm}$ as the $\zeta^{\pm 1}$-eigenspaces of $r$ on fermions.
Then, $\mathcal{F}_-$ consists of fermions whose scalar partner is $r$-invariant, and $\mathcal{F}_+$ consists of those whose scalar partner has eigenvalue $\zeta^2$, together with the $B$ gauginos.
For the case that $r$ is a pure $\ZNR{M}$ element with $\xi = \zeta^q$, the fermion has $R$-charge $f = q-1$. 
$\mathcal{F}_-$ then consists of the fermions of $R$-charge $0$ multiplets, and $\mathcal{F}_+$ consists of the fermions of $R$-charge~$2$ multiplets plus all gauginos.

We now state the following hypotheses we will make use of throughout our argument:
\begin{assumptions}
	\item\label{ass:comp_conn} $G$ is compact and connected, with $G_0$, $T_0$, and $\Delta_0$ as defined above.
	\item\label{ass:r_micro} $r$ is a microscopic symmetry, i.e.\  it acts on the full field content, not just light fields, and is linear, unitary, unbroken, and satisfies $\zeta^2 \neq 1$.
	\item\label{ass:r_comm} $r$ commutes with a closed connected subgroup $B$ with $G_0 \subseteq B \subseteq G$.
  The direct product case where $r$ commutes with $G$ is $B = G$. 
  Commuting with $B$ excludes outer-automorphism actions.
	\item\label{ass:vac} By definition, the vacuum is invariant under $\Sigma_0$, with $G_0 \subseteq \Sigma_0$ and $r \in \Sigma_0$. 
	Therefore, the fermion mass bilinear $m$, a symmetric bilinear form on the fermion space $\mathcal{F}$, is $G_0$- and $r$-invariant.
	\item\label{ass:finitely_many_multiplets} There are finitely many fields, so $\mathcal{F}$ is a finite-dimensional unitary $G$-representation, i.e.\ chiral fermions in $G$-multiplets plus gauginos in $\ad G$.
\end{assumptions}

\subsection{Intermediate Results}

To proceed to a statement of the no-go theorem and its proof, we first note a few intermediate results.

Here, we make use of our earlier observation that since $B$ and $r$ commute, they share a simultaneous eigenspace.
The fermion space $\mathcal{F}$ splits into $r$-eigenspaces, which we label as $\mathcal{F}_\lambda$.
Since $r$ is unitary and commutes with $B$, each $\mathcal{F}_\lambda$ is also a $B$-representation.
It is useful to also construct the spaces $\mathcal{F}_\lambda(\chi)$ in the following manner.
Recall that since $G_0 \subseteq B$, $B$ contains the maximal torus of $G_0$, i.e.\ $T_0$.
We can then take the spaces $\mathcal{F}_\lambda$, which are necessarily $B$-representations, and split them further into $T_0$-weight spaces.
We denote the $\mathcal{F}_\lambda$ space with $T_0$-weight $\chi$ as $\mathcal{F}_\lambda(\chi)$.
That is, $r$ acts on states in $\mathcal{F}_\lambda(\chi)$ by multiplying by $\lambda$, while $t \in T_0$ acts on states by the phase $\chi(t)$. 

Invariance of the vacuum under $\Sigma_0$, \Cref{ass:vac}, requires $m(r\psi_a,r\psi_b) = m(\psi_a,\psi_b)$, i.e.\ $m(\psi_a,\psi_b) = \lambda_a \lambda_b m(\psi_a,\psi_b)$ for states $\psi_a \in \mathcal{F}_{\lambda_a}$ and $\psi_b \in \mathcal{F}_{\lambda_b}$.
Equality of both sides implies that $m$ pairs $\mathcal{F}_\lambda$ only with $\mathcal{F}_{\lambda^{-1}}$. 
By \Cref{ass:r_micro}, $\zeta^{-1} \neq \zeta$, so $m$ vanishes on $\mathcal{F}_\pm \times \mathcal{F}_\pm$, and pairs $\mathcal{F}_-$ only with $\mathcal{F}_+$.
In particular, gaugino Majorana masses, which would live in $\mathcal{F}_+ \times \mathcal{F}_+$, are forbidden.
On $\mathcal{F}_- \oplus \mathcal{F}_+$ the mass matrix is given by
\begin{equation}\label{eq:offdiag}
	\begin{pmatrix}
		0 & \mathfrak{M}\\
		\mathfrak{M}^{\mathsf{T}} & 0		
	\end{pmatrix}\;,\quad \mathfrak{M}:\mathcal{F}_- \to \mathcal{F}_+^\vee\;, \quad \mathfrak{M}(\psi) = m(\psi,\cdot)\;,
\end{equation}
where $\mathcal{F}_+^\vee$ denotes the dual space to $\mathcal{F}_+$.
This sector decouples from the other sectors, and its zero modes are $\ker \mathfrak{M} \oplus \ker \mathfrak{M}^{\mathsf{T}}$ with $\dim \ker \mathfrak{M}^{\mathsf{T}} = \dim \coker \mathfrak{M}$.

Restricted to $T_0 \subseteq G_0$,  $\mathcal{F}_\pm$ and $\mathcal{F}_\pm^\vee$ split into $T_0$-weight spaces as discussed earlier.
$T_0$-invariance of $m$ says that $m\bigl(\mathcal{F}_-(\chi),\mathcal{F}_+(\chi')\bigr) = 0$ unless $\chi+\chi' = 0$.
Therefore, $\mathfrak{M}$ decomposes in blocks as $\mathfrak{M} = \oplus_\chi \mathfrak{M}_\chi$ with
\begin{equation}
	\mathfrak{M}_\chi : \mathcal{F}_-(\chi) \to \mathcal{F}_+(-\chi)^\vee\;.
\end{equation}
We define $L_-(\chi) = \dim \ker \mathfrak{M}_\chi$, i.e.\ the number of massless $\zeta^{-1}$-eigenspace fermions at weight $\chi$, $L_+(-\chi) = \dim \coker \mathfrak{M}_\chi$, i.e.\ the number of massless $\zeta^{+1}$-eigenspace fermions at weight $-\chi$, and
\begin{equation}
	\ind \mathfrak{M}_\chi = L_-(\chi) - L_+(-\chi)\;.
\end{equation}
For an $R$-symmetry such as \ZNR{M} or $\U{1}_R$ with superpotential $R$-charge $2$, $L_-(\chi)$ counts the massless fermions of weight $\chi$ from $R$-charge $0$ multiplets, whereas $L_+(-\chi)$ counts the massless fermions of weight $-\chi$ from $R$-charge $2$ multiplets together with the massless gauginos of weight $-\chi$.
By rank-nullity, with $s = \rank \mathfrak{M}_\chi$, we have
\begin{equation}\label{eq:ind_F}
	\ind \mathfrak{M}_\chi = \bigl(\dim \mathcal{F}_-(\chi) - s\bigr) - \bigl(\dim \mathcal{F}_+(-\chi) - s\bigr) = \dim \mathcal{F}_-(\chi) - \dim \mathcal{F}_+(-\chi)\;.
\end{equation}
Note that this corresponds to the cancellation of massive pairs in the arguments of \cite{Goodman:1985bw}.

Now let $V$ be a finite-dimensional $B$-representation, $n \in N_{B}(T_{0})$, and $V(\chi)\defeq\bigl\{v \in V \,\big\vert\, t\,v = \chi(t)\,v\;\forall t \in T_{0}\bigr\}$.
Then, for $v \in V(\chi)$,
\begin{equation}
	t\,(n\,v) = n\,(n^{-1}\,t\,n)\,v = \chi(n^{-1}\,t\,n)\,n\,v = \bigl(w \cdot \chi\bigr)(t) \,n\,v \;,
\end{equation}
where the second equality uses that $n^{-1}\, t\, n \in T_0$ for $n \in N_B(T_0)$, by definition.
The third equality rewrites the conjugation of $t$ by $n$ as the action of an element $w \in \Weyl(B,T_0)$ on the weight $\chi$.
We can then conclude that $n\,V(\chi) \subseteq V(w \cdot \chi)$.
Repeating the argument for $n^{-1}$ shows $n^{-1}\,V(w\cdot \chi) \subseteq V(\chi)$, so that $n\,V(\chi) = V(w\cdot \chi)$ and $\dim V(\chi) = \dim V(w\cdot \chi)$.
Applying this to $V = \mathcal{F}_{\pm}$ and using~\eqref{eq:ind_F}, we find
\begin{equation}\label{eq:ind_weights}
	\ind \mathfrak{M}_{w\cdot \chi} = \ind \mathfrak{M}_{\chi}\quad \forall w \in \Weyl(B,T_{0})\;.
\end{equation}
Note that $\mathfrak{M}$ itself is only $G_{0}$-equivariant, and the invariance above comes from~\eqref{eq:ind_F}, which involves only the dimensions of $B$-representations.

\subsection{The No-Go Theorem and Its Proof}

\begin{theorem}[Abstract no-go theorem]\label{thm:abstract}
	Assume \ref{ass:comp_conn}--\ref{ass:finitely_many_multiplets} within a \ac{4D} $\mathcal{N}=1$ \ac{GUT}. 
  Let $\beta \in \Delta_{0}$ and $w \in \Weyl(B,T_{0})$ with $w\beta \not\in \Delta_{0}$ (so $\beta$, $w\beta \neq 0$). 
  Suppose that for $\chi \in \{\beta,w\beta\}$
	\begin{enumerate}[label=(\alph*),ref=(\alph*),leftmargin=*]
		\item $L_{-}(\chi) = 0$, i.e.\ no massless $\zeta^{-1}$-eigenspace fermion has weight $\beta$ or $w\beta$,
		\item $L_{+}(-\chi) = \dim (\mathfrak{g}_{0} \otimes \mathds{C})_{-\chi}$, i.e.\ the massless $\zeta^{+1}$-eigenspace fermions at $-\beta$ and $-w\beta$ are exactly the $G_{0}$ gauginos there, so no massless chiral $\zeta^{+1}$-eigenspace fermion, and no massless broken-generator gaugino,
	\end{enumerate}
	then the index identity~\eqref{eq:ind_weights} is violated, i.e.\ such a massless spectrum cannot occur. Equivalently, there must be additional massless fermions at weight $\pm\beta$ or $\pm w\beta$.
\end{theorem}

\begin{proof}
  Since $\Delta_{0} = -\Delta_{0}$ and root spaces are one-dimensional, $\dim(\mathfrak{g}_{0}\otimes \mathds{C})_{-\beta} = 1$, and $\dim(\mathfrak{g}_{0}\otimes \mathds{C})_{-w\beta} = 0$ because $-w\beta \not\in \Delta_{0}$, as $w\beta \not\in \Delta_{0}$ and $\Delta_{0} = -\Delta_{0}$.
  We also have that $-w\beta \neq 0$.
  Then,
  \begin{equation}
    \ind \mathfrak{M}_{\beta} = 0 - 1 = -1\;,\quad \ind \mathfrak{M}_{w\beta} = 0 - 0 = 0\;,
  \end{equation}
  contradicting~\eqref{eq:ind_weights}.
\end{proof}
The Lean formalization of this proof is summarized in \Cref{sec:lean}, with the relevant code presented in \Cref{app:Lean}.

\subsection{Implications of the Theorem}

In the following sections, we will apply this theorem to \ac{4D} $\mathcal{N} = 1$ \ac{GUT} models.
In this process, we will first identify the data $G$, $G_{0}$, $T_{0}$, $\Delta_{0}$, $r$, $B$, and $\Weyl(B,T_{0})$.
Then, we verify that the hypotheses \ref{ass:comp_conn}--\ref{ass:finitely_many_multiplets} are satisfied.
Next, we identify the relevant $\mathcal{F}_{\pm}$ sectors, and determine whether there are $\beta$ and $w$ that we can use to derive the contradiction.
Such a pair exists if and only if $\Delta_{0}$ is not stable under $\Weyl(B,T_{0})$.
In words, this means there exists some element of the Weyl group that preserves the torus of $G_{0}$ and maps an unbroken root to a weight that is a broken root.
In practice, we can construct $\Weyl(B,T_{0})$ using \Cref{eq:weylBT0} and then test the stability of $\Delta_0$ under $\Weyl(B,T_{0})$.
We will see that the theorem of \cite{Fallbacher:2011xg} follows. 
We will also discuss, using key insights from \cite{Goodman:1985bw}, how higher-dimensional models of grand unification evade this theorem.

\section{Application of the No-Go Theorem to \texorpdfstring{$\ZNR{M}$}{Z\_M\^{}R} Symmetries in \texorpdfstring{\SU{5}}{SU(5)} GUTs}\label{sec:SU5}

Let us consider a \ac{4D} $\mathcal{N}=1$ supersymmetric \SU{5} \ac{GUT}.
We assume that there are finitely many multiplets, cf.\ \Cref{ass:finitely_many_multiplets}.
Suppose that the vacuum leaves \GSM{} unbroken, and also an element $r_R$ of an unbroken \ZNR{M} symmetry, where the action of $r_R$ on the superspace coordinates is the phase $\zeta = \ee^{2\piup\ii/M}$.
In particular, $\zeta^{2} \neq 1$, so $M\not\in \{1,2\}$.\footnote{%
Note that there are no $\ZNR{1}$ or $\ZNR{2}$ symmetries. 
A $\ZNR{1}$ is the trivial group. 
Under a $\ZNR{2}$ the superpotential is invariant, and the only difference from an ordinary $\Z{2}$ symmetry is that the superspace coordinates transform as $\theta\mapsto-\theta$. 
However, every relativistic quantum field theory possesses the symmetry $(-1)^{F}$, a rotation by $2\piup$, under which all fermions change sign, and in superspace it acts precisely as $\theta\mapsto-\theta$ on the superspace coordinate $\theta$. 
Multiplying a $\ZNR{2}$ by $(-1)^{F}$ therefore yields an ordinary $\Z{2}$ symmetry, which is physically equivalent. 
This is the same equivalence that relates $R$ parity \cite{Fayet:1974jb} and matter parity (or family reflection symmetry \cite{Dimopoulos:1981zb}) in the \ac{MSSM}.\label{ftn:no_order_1_or_2_R-symmetries}}
The inputs to the abstract no-go theorem are $G = B = \SU{5}$ and $G_{0} = \GSM$. 
$T_{0}$ is the diagonal torus and is maximal in both $G_{0}$ and $G = B$, so $\Weyl(B,T_{0}) = S_{5}$, the Weyl group of \SU{5}.
Let us further denote the fundamental weights by $e_i$. 
Roots are then given as differences of weights, $e_i-e_j$.
$\Delta_{0}$ are the roots of $G_{0} = \GSM$, which with color indices $\{0,1,2\}$ and weak indices $\{3,4\}$ corresponds to the roots $e_{i} - e_{j}$ of \SU{5} where both $i,j$ are in $\{0,1,2\}$ or $\{3,4\}$.
There are 8 unbroken roots, and the other 12 correspond to the roots for the heavy leptoquark gauge bosons $V$ of \SU{5}.
By assumption, the \ZNR{M} is unbroken by the vacuum, so the vacuum is invariant under the action of both $G_{0} = \GSM$ and $r_R$.
One can then verify that all of this input data satisfies the assumptions \ref{ass:comp_conn}--\ref{ass:finitely_many_multiplets}.

Now we identify a pair of roots, $\{\beta,w\,\beta\}$, that will lead to a contradiction. 
The element $w = (1\, 3)$ of $S_{5}$ sends the gluon root $\beta = e_{0}-e_{1}$ to the $V$ root $w\beta = e_{0} - e_{3}$, with $\beta \in \Delta_{0}$ and $w\beta \not\in \Delta_{0}$.
Next, we input a light chiral spectrum and check if such a spectrum is inconsistent with the existence of the unbroken \ZNR{M} in the \ac{GUT}.
Let us assume the light spectrum is the \ac{MSSM}, possibly with additional gauge singlets. 
Recall that $L_{-}(\chi)$ with $\chi \in \{\beta,w\,\beta\}$ denotes the number of massless fermions of weight $\chi$ and $R$-charge $-1$.
The gauginos transform under $r_R$ in the same way as the $\theta$ superspace coordinates, so they transform with $\zeta^{+1}$ and are not in the $\zeta^{-1}$-eigenspace.
There are no other massless chiral fermions in the spectrum that can sit at a root, so $L_{-}(\chi) = 0$ regardless of the $R$-charge assignments to the \ac{MSSM} fields and singlets.
Similarly, for $L_{+}(-\chi)$ the only massless fermions that appear at a root in the $\zeta^{+1}$-eigenspace are the gauginos of \GSM.
At $\beta = e_{0} - e_{1}$, there is a gluino and $L_{+}(-\beta) = L_{+}(\beta) = 1$.
At $w \beta = e_{0} - e_{3}$, there is no massless fermion since the $V$ gauge bosons are heavy, so $L_{+}(-w\beta) = L_{+}(w\beta) = 0$.
This then provides the contradiction,
\begin{equation}\label{eq:contradiction}
  \left.\begin{aligned}
\ind \mathfrak{M}_{e_{0} - e_{1}} &= -1\\ 
\ind \mathfrak{M}_{e_{0} - e_{3}} &= 0
 \end{aligned}\right\}
	\; \implies  \ind \mathfrak{M}_{e_{0} - e_{1}} \neq \ind \mathfrak{M}_{e_{0} - e_{3}}\;.
\end{equation}
This means that the spectrum cannot be the exact \ac{MSSM} with additional gauge singlets.

The extension to non-standard embeddings is straightforward.
The choice of embedding changes the position of $\mathfrak{t}_{0}$ in the Cartan of $G$, which fixes $\Weyl(G,T_{0})$.
The no-go theorem can then be applied if $\Delta_{0}$ is not stable under $\Weyl(G,T_{0})$.
This fails, for example, for flipped $\SU{5} \times \U{1}_{X}$, where $\Weyl(G,T_{0})$ shrinks to $\Weyl(\GSM)$, under which $\Delta_{0}$ is stable.

\subsection{Comparison with the Original Proof}
\label{sec:gap}

The general case in~\cite[Section~2.1.2]{Fallbacher:2011xg} proceeds in two steps.
First, all pairs of representations that admit an $\SU5\times\ZNR{M}$-invariant mass term are removed from the fields of $R$-charge $0$ and $R$-charge $2$.
Such a pair consists of a representation $\rep r$ of $R$-charge $0$ and its conjugate $\crep r$ of $R$-charge $2$, for which $\rep r\,\crep r$ is an allowed mass term.
The removal is decided by group theory alone.
It asks whether a mass term is allowed, not whether the fields are actually massive in the vacuum.
Second, the highest weights of the remaining representations are compared.
After the removal, the two sectors no longer share an irreducible representation, so their highest weights differ.
At the larger of the two, say the highest weight $\Lambda_0$ of the fields of $R$-charge $0$, there are states with no partner of $R$-charge $2$ at the opposite weight.
These states, and their Weyl images, therefore remain massless.
The case analysis of~\cite{Fallbacher:2011xg} then shows that they cannot all be harmless, and give rise to light exotics.
The case $\Lambda_2>\Lambda_0$ is analogous.
This comparison uses that the remaining fields of $R$-charge $0$ still contain the vacuum, since otherwise the heavy gauge bosons could not acquire masses.\footnote{Specifically, the case $\Lambda_0=0$, in which only $\SU5$ singlets of $R$-charge $0$ remain, is excluded in this way.
A singlet \ac{VEV} is invariant under all of \SU5, so it cannot give masses to the heavy gauge bosons~$V$.}
However, the first step may remove the multiplet that carries the vacuum.
Consider, for instance, fields of $R$-charge $0$ and fields of $R$-charge $2$ that both transform as $\rep{24}\oplus\rep{75}$.
All of these representations are real, so all of them can be removed in pairs.
In this case, no highest weights remain, and the argument does not apply.
The conclusion of~\cite{Fallbacher:2011xg} is nevertheless correct.
Removing invariant pairs does not change the formal difference of the two sectors.
By this we mean, for every weight $\chi$, the number of states of $R$-charge $0$ at $\chi$ minus the number of states of $R$-charge $2$ at $-\chi$.
A removed pair $(\rep r,\crep r)$ contributes equally to both numbers.
It is therefore legitimate for the comparison of highest weights.
It is not legitimate for the vacuum and its Goldstone directions, which depend on the full field content.
Where the \ac{VEV} sits, which directions are eaten, and which states are actually massless depend on the superpotential of the full theory. 
Once the \ac{VEV} is switched on, the removed fields mix with the field that carries it, with the Goldstone modes, and with the gauginos. 

On the other hand, the proof of \Cref{thm:abstract} does not remove any pairs.
The index~\eqref{eq:ind_F} is computed for the full field content, and the vacuum enters only through the gaugino count~(b).
Hence the example above is covered by the \SU5 result~\eqref{eq:contradiction}.
In particular, in this example the two sectors have the same content.
Therefore, the chiral multiplets cancel in~\eqref{eq:ind_F} and only the gaugino remains.
The index is therefore $-1$ at all twenty roots.
An exact \ac{MSSM} would instead require $-1$ at the unbroken roots and $0$ at the broken ones.
The special case in which only $\rep{24}$-plets are present is treated separately in~\cite[Section~2.1.1]{Fallbacher:2011xg}, and its proof is correct.

\section{\texorpdfstring{$\ZNR{M}$}{Z\_M\^{}R} Symmetries with Larger Gauge Groups}\label{sec:Larger}

Now we turn to the application of the theorem to \acp{GUT} with larger gauge groups $G$ broken to $G_0 = \GSM$.

\begin{theorem}
	Let $G$ be compact and connected, and let $\SU5 \subseteq G$ be a subgroup with $\GSM \subset \SU5$ embedded in the standard way, i.e.\ $\SU3 \times \SU2$ is the block subgroup and hypercharge $\Generator{Y} \in \mathfrak{su}(5)$.
	We assume \ref{ass:comp_conn}--\ref{ass:finitely_many_multiplets}, with $G_0 = \GSM$.
	Then, if an unbroken symmetry element $r \in \Sigma_0$ with $\zeta^2 \neq 1$ commutes with \SU5, the massless spectrum cannot be the \ac{MSSM} plus gauge singlets.
\end{theorem}

\begin{proof}
	Choose $B = \SU5$.
	Since \SU5 is closed, connected, contains \GSM, and commutes with $r \in \Sigma_0$ by assumption, this satisfies \Cref{ass:r_comm}.
	$T_0$ is the maximal torus of $G_0 = \GSM$, and with the standard embedding $T_0$ is also the maximal torus of $B = \SU5$.
	As a result, $\Weyl(B,T_0) = S_5$.
	By \Cref{ass:finitely_many_multiplets}, the fermion space $\mathcal{F}$ is a $G$-representation, which when restricted to \SU5 must be an \SU5-representation as every $G$-representation branches to an \SU5-representation.
	Furthermore, $r$ commutes with \SU5, so each $\mathcal{F}_\pm$ is also an \SU5-representation.
	The index relation \Cref{eq:ind_weights} then holds for \SU5, and as before we can take $\beta = e_0 - e_1 \in \Delta_0$ and $w = (1\, 3)$ so that $w \beta = e_0 - e_3 \not\in \Delta_0$.
	Now we verify the requirements (a) and (b) of \Cref{thm:abstract}.
	
	We first verify (a), namely $L_-(\chi) = 0$ for $\chi \in \{\beta,w\beta\}$.
	As in the discussion for the $G = \SU5$ case, the \ac{MSSM} contains no massless $\zeta^{-1}$ fields sitting at a root.
	The singlets sit at weight $0$, which is not a root, and the gauginos are in $\mathcal{F}_+$, not $\mathcal{F}_-$.
	Hence, no massless fermion sits at $\beta$ or $w\beta$, and we conclude $L_-(\chi) = 0$.
	
	Now we verify (b), that $L_+(-\chi) = \dim(\mathfrak{g}_0 \otimes \mathds{C})_{-\chi}$.
	The above argument for (a) establishes that there are no massless chiral $\zeta^{+1}$-eigenspace fermions at $-\beta$ and $-w\beta$.
	At $-\beta$, the gluino is massless since $G_0 = \GSM$ is unbroken, and it lies in $\mathcal{F}_+$ as a gaugino.
	At $-w\beta$, there is a massive gaugino since every gauge generator of $G$ at this weight lies outside of $\mathfrak{g}_0 = \mathfrak{g}_{\mathrm{SM}}$, and is broken.
	Then the proof proceeds as in the \SU5 case, where there is a massless gluino at weight $\beta = e_0 - e_1$, and a massive gaugino outside of $\GSM$ at weight $w\beta = e_0 - e_3$.
	As a result, $\ind \mathfrak{M}_{e_0 - e_1} = -1$, and $\ind \mathfrak{M}_{e_0 - e_3} = 0$.
	This provides the contradiction that $\ind \mathfrak{M}_{e_0 - e_1} \neq \ind \mathfrak{M}_{e_0 - e_3}$, so the spectrum cannot be the exact \ac{MSSM} with additional gauge singlets.
\end{proof}

\section{How Higher-Dimensional Models of Grand Unification \texorpdfstring{Evade}{Evade} the No-Go Theorem}\label{sec:extra_dim}

In order to understand how higher-dimensional models evade the no-go theorem, let us first recall, at a qualitative level, why the no-go theorem applies.
The basic argument can be illustrated in an \SU{5} model with an arbitrary $R$-symmetry. 
Assume that a \ac{GUT} Higgs $H$ transforming in the adjoint representation breaks \SU{5} to \GSM.
Under \GSM, the adjoint decomposes as
\begin{equation}\label{eq:24_decomposition}
 \rep{24}\to(\rep{8},\rep{1})_{0}\oplus(\rep{1},\rep{3})_{0}\oplus(\rep{1},\rep{1})_{0}\oplus(\rep{3},\rep{2})_{-5/6}\oplus(\crep{3},\rep{2})_{5/6}\;.
\end{equation}
In order to preserve the $R$-symmetry, $H$ has to carry $R$-charge $0$.
This implies that any superpotential term built from $H$ alone has $R$-charge $0\ne2$ and is, therefore, forbidden.
The components $(\rep{3},\rep{2})_{-5/6}\oplus(\crep{3},\rep{2})_{5/6}$ are eaten by the heavy gauge bosons of $\SU{5}/\GSM$.
On the other hand, the remaining components $(\rep{8},\rep{1})_{0}$, $(\rep{1},\rep{3})_{0}$ and $(\rep{1},\rep{1})_{0}$ remain massless. 
These are light, exotic states that go beyond the \ac{MSSM}.

One may now try to introduce mass partners for the exotics. 
In \cite{Fallbacher:2011xg}, it has been shown that these mass partners have to be in the same congruency class as the adjoint. 
For instance, introducing an additional $\rep{24}$-plet $H'$ with $R$-charge $2$ does not help. 
The mass term $H\,H'$ is allowed, and allows us to pair up the $(\rep{8},\rep{1})_{0}$, $(\rep{1},\rep{3})_{0}$ and $(\rep{1},\rep{1})_{0}$ components of $H$ and $H'$.
However, the $(\rep{3},\rep{2})_{-5/6}\oplus(\crep{3},\rep{2})_{5/6}$ components of $H'$ have no partners since the corresponding components of $H$ are eaten. 
We are thus, again, left with massless exotics. 

More generally, let us consider $n_0$ adjoints with $R$-charge $0$ and $n_2$ adjoints with $R$-charge $2$.
The absence of massless $(\rep{8},\rep{1})_{0}$ states requires $n_0=n_2$, while the absence of massless $(\rep{3},\rep{2})_{-5/6}$ states requires $n_0-1=n_2$. 
These conditions are incompatible with each other for any finite numbers $n_0$ and $n_2$. 
This counting is the special case of~\eqref{eq:ind_weights} in which only adjoints are present.

The basic argument can be generalized to arbitrary \SU{5} representations, to larger \ac{GUT} groups $G\supset\SU{5}$ and to singlet extensions of the \ac{MSSM}~\cite{Fallbacher:2011xg}.
The general version rests on three assumptions.
First, the \ac{GUT} is four-dimensional and based on a gauge group $G\supset\SU{5}$.
Second, the \ac{GUT} symmetry is broken spontaneously.
Third, there are only finitely many fields.
Under these assumptions it is impossible to obtain a low-energy effective theory that has both the exact \ac{MSSM} field content and a residual $R$-symmetry. 
See \Cref{sec:gap} for additional details.
In \Cref{sec:No-Go_Theorem}, we have turned these arguments into a precise, formalized theorem. 

The argument, and the theorem, no longer hold if there are infinitely many $\rep{24}$-plets, or other appropriate representations.
As a consequence, higher-dimensional models of grand unification evade the no-go theorem.  
This is because, from a \ac{4D} perspective, they contain an infinite number of multiplets~\cite{Goodman:1985bw}, thus violating \Cref{ass:finitely_many_multiplets}. 
In particular, scenarios in which the \ac{GUT} symmetry gets broken by compactification evade the no-go theorem. 
These scenarios include the so-called orbifold \acp{GUT} and orbifold compactifications of the heterotic string~\cite{Dixon:1985jw,Dixon:1986jc} (see e.g.\ \cite{Quiros:2003gg,Ramos-Sanchez:2024keh} for more references).

\section{Modular Symmetries}\label{sec:modular}

\subsection{Modular Symmetries \texorpdfstring{Are}{Are} $R$-Symmetries}

We start by pointing out that in a supersymmetric theory that also describes gravity, modular symmetries are $R$-symmetries. 
Under a modular transformation $\gamma\defeq\begin{psmallmatrix}
 a & b\\ c & d  
\end{psmallmatrix}\in\Group{SL}{2,\mathds{Z}}$, the modulus $\ModulusU$, chiral superfields $\Phi^{(i)}$ and superpotential $\mathscr{W}$ transform as 
\begin{subequations}\label{eq:modular_trafos}
\begin{align}
 \ModulusU&\xmapsto{~\gamma~}\frac{a\,\ModulusU+b}{c\,\ModulusU+d}\;,\label{eq:modular_trafo_U}\\ 
 \Phi^{(i)}&\xmapsto{~\gamma~}\rho_{\rep{r}^{(i)}}(\gamma)\,(c\,\ModulusU+d)^{\ModularWeight^{(i)}}\,\Phi^{(i)}\eqdef \rho_{\rep{r}^{(i)}}(\gamma)\,j^{\ModularWeight^{(i)}}\,\Phi^{(i)}\;,\label{eq:modular_trafo_Phi}\\
 \mathscr{W}&\xmapsto{~\gamma~}(c\,\ModulusU+d)^{\ModularWeight_{\mathscr{W}}}\,\mathscr{W}\eqdef j^{\ModularWeight_{\mathscr{W}}}\,\mathscr{W}\;.\label{eq:modular_trafo_W}
\end{align}    
\end{subequations}
Here, we introduced the shorthand $j\defeq (c\,\ModulusU+d)$, and $\ModulusU$ denotes the modulus and $\Phi^{(i)}$ the chiral superfields of the theory.  
$\ModularWeight^{(i)}$ and $\ModularWeight_{\mathscr{W}}$ stand for the modular weights of the chiral superfields and the superpotential, respectively. 
With this sign convention, a coupling $Y(\ModulusU)\prod_i\Phi^{(i)}$ requires $Y\in\Mod{\ModularWeight_{\mathscr{W}}-\sum_i \ModularWeight^{(i)}}$.
Here, $\Mod{k}$ denotes the space of holomorphic modular forms of weight $k$ (regular at the cusp) for the full modular group, unless specified otherwise~\cite{Liu:2021gwa}. 
The modular group is generated by 
\begin{subequations}
\begin{align}
  \Generator{S}&\defeq\begin{pmatrix*}[r]
  0 & 1   \\ -1 &0 
\end{pmatrix*}\;,\label{eq:Generator_S_def}\\ 
  \Generator{T}&\defeq\begin{pmatrix}
 1 & 1\\ 0 & 1 
\end{pmatrix}\;.\label{eq:Generator_T_def}
\end{align}  
\end{subequations}

To understand why modular symmetries are necessarily $R$-symmetries, recall that the unique supersymmetric framework that incorporates gravity is \ac{SUGRA}.
As is well known, in \ac{SUGRA} the so-called $\mathscr{G}$-function is given by
\begin{equation}\label{eq:G-function}
	\mathscr{G}=K+\ln\lvert\mathscr{W}\rvert^2\;.
\end{equation}
A typical K\"ahler potential is given by
\begin{equation}\label{eq:Kaehler_one_plane}
	K_\mathrm{modulus}=-n_\mathrm{tori}\,\ln\bigl(-\ii\ModulusU+\ii\overline{\ModulusU}\bigr)\;,
\end{equation}
where $n_\mathrm{tori}$ denotes the number of tori $\ModulusU$ parametrizes. 
Since $K_\mathrm{modulus}$ transforms nontrivially under \eqref{eq:modular_trafo_U}, the superpotential has to undergo the transformation \eqref{eq:modular_trafo_W} to ensure modular invariance of the $\mathscr{G}$-function \eqref{eq:G-function}. 
Indeed, $\im(\gamma\,\ModulusU)=\im\ModulusU/\lvert c\,\ModulusU+d\rvert^2$, hence $K_{\mathrm{modulus}}(\gamma\ModulusU)=K_{\mathrm{modulus}}(\ModulusU)+n_{\mathrm{tori}}\ln\lvert c\,\ModulusU+d\rvert^2$. 
Invariance of $\mathscr{G}$ then requires $\mathscr{W}\mapsto(c\,\ModulusU+d)^{-n_{\mathrm{tori}}}\mathscr{W}$, up to a possible multiplier. 
Specifically, for the K\"ahler potential \eqref{eq:Kaehler_one_plane} one has 
\begin{equation}\label{eq:typical_k_W}
 \ModularWeight_{\mathscr{W}}=-n_\mathrm{tori}
\end{equation}
for the modular transformations \eqref{eq:modular_trafos}. 
In particular, in the presence of gravity, modular invariance and \ac{SUSY} imply that modular symmetries are $R$-symmetries.

In heterotic orbifolds, discrete $R$-symmetries have traditionally been linked to the H-momentum selection rule \cite{Hamidi:1986vh,Dixon:1986qv,Kobayashi:2004ya,CaboBizet:2013hms,Nilles:2013lda}. 
However, as is well known, in string compactifications $R$-symmetries really originate from modular symmetries~\cite{Dixon:1989fj,Ibanez:1992uh}. 
This is, of course, in agreement with our earlier statement that in a supersymmetric framework that describes gravity, modular symmetries are $R$-symmetries. 
It is, therefore, natural to ask what the implications of the no-go theorems for $R$-symmetries are for modular symmetries.

\subsection{\texorpdfstring{$R$}{R}-Symmetries as Stabilizers}
\label{sec:stabilizers}

There are certain critical values of $\ModulusU$, $\ModulusU_0$, which are invariant under a subgroup of $\SL$. 
The modular transformations $\gamma$ with $\gamma\,\ModulusU_{0}=\ModulusU_{0}$ form the so-called stabilizer subgroup $\Gamma_{\langle\ModulusU\rangle}$ of \SL. 
For $\gamma_0\in\Gamma_{\langle\ModulusU\rangle}$ one then has
\begin{subequations}\label{eq:modular_trafos_stabilizer}
\begin{align}
 \ModulusU&\xmapsto{~\gamma_0~}\ModulusU\;,\label{eq:modular_trafo_U_0}\\ 
 \Phi^{(i)}&\xmapsto{~\gamma_0~}\rho_{\rep{r}^{(i)}}(\gamma_0)\,j_0^{\ModularWeight^{(i)}}\,\Phi^{(i)}\;,\label{eq:modular_trafo_Phi_0}\\
 \mathscr{W}&\xmapsto{~\gamma_0~}j_0^{\ModularWeight_{\mathscr{W}}}\,\mathscr{W}\;,\label{eq:modular_trafo_W_0}
\end{align}    
\end{subequations}
where now $j_0\defeq c\,\ModulusU_0+d$.
Provided that $j_0^{\ModularWeight_{\mathscr{W}}}\ne1$, \eqref{eq:modular_trafos_stabilizer} is an $R$-symmetry. 
In \Cref{tab:stabilizers} we list the nontrivial stabilizers of $\SL$. 
Note that different modular symmetries such as congruence subgroups of $\SL$ can have different stabilizers~\cite{diamond2005first}.

\begin{table}[htb]
  \centering
  \setlength{\tabcolsep}{4pt}
  \begin{tabular}{lllllll}
  \toprule
  $\ModulusU_{0}$ & generator & $j$ & superpotential phase & \multicolumn{3}{c}{linearly realized $\ZNR{M}$}\\
  & & & & $n_\mathrm{tori}=1$ & $n_\mathrm{tori}$ odd & $n_\mathrm{tori}$ even\\
  \midrule
  generic & $\Generator{S}^2=-\mathds{1}$ & $-1$ & $(-1)^{\ModularWeight_{\mathscr{W}}}$ & $\ZNR{4}$ & $\ZNR{4}$ & ---\\
  $\ii$ & $\Generator{S}$ & $-\ii$ & $(-\ii)^{\ModularWeight_{\mathscr{W}}}$ & $\ZNR{8}$ & $\ZNR{8}$ & \begin{tabular}[t]{@{}l@{}}$\ZNR{4}$ for $n_\mathrm{tori}=2\bmod4$\\ --- for $4\mid n_\mathrm{tori}$\end{tabular}\\
  $\omegaup=\ee^{2\piup\ii/3}$ & $-\Generator{S}\,\Generator{T}$ & $\ee^{\ii\piup/3}$ & $\ee^{\ii\piup\ModularWeight_{\mathscr{W}}/3}$ & $\ZNR{12}$ & \begin{tabular}[t]{@{}l@{}}$\ZNR{12}$ for $3\nmid n_\mathrm{tori}$\\ $\ZNR{4}$ for $3\mid n_\mathrm{tori}$\end{tabular} & \begin{tabular}[t]{@{}l@{}}$\ZNR{6}$ for $3\nmid n_\mathrm{tori}$\\ --- for $3\mid n_\mathrm{tori}$\end{tabular}\\
  $\ii\infty$ & $\Generator{T}$ & $1$ & $1$ & $\ZNR{4}$ & $\ZNR{4}$ & ---\\
  \bottomrule
  \end{tabular}
  \caption{Stabilizers of the vacuum value $\ModulusU_{0}$ in $\SL$ and the corresponding superpotential phases.
  ``---'' indicates that there is no linearly realized symmetry. The last three columns refer to the full stabilizer, which always contains the center $-\mathds{1}$.}
  \label{tab:stabilizers}
\end{table}

The stabilizer symmetries at $\ii$, $\omegaup$ and $\ii\infty$ in \Cref{tab:stabilizers} act as \ZNR{M} symmetries unless $n_\mathrm{tori}$ takes special values. 
In particular, in the arguably most interesting case of a single torus, i.e.\ $n_\mathrm{tori}=1$, there are linearly realized $\ZNR{4}$, $\ZNR{8}$ and $\ZNR{12}$ symmetries at generic $\ModulusU$, at $\ModulusU=\ii$ and at $\ModulusU=\omegaup$, respectively. 
The superpotential has $R$-charge $2\ModularWeight_{\mathscr{W}}$. 
We introduced the factor 2 here to make sure that the superspace coordinates $\theta$ have integer $R$-charge.\footnote{In string compactifications, the modular weights are often fractional~\cite{Ibanez:1992hc}. In this case, additional rescaling is necessary to make the discrete $R$-charges integer.}

\subsection{\texorpdfstring{The Center of \SL{} and a Modular \texorpdfstring{$\ZNR{4}$}{Z4R} Symmetry}{The Center of SL(2,Z) and a Modular Z4R Symmetry}}
\label{sec:center_and_Z_4_r}

At a generic point in $\ModulusU$ moduli space, the stabilizer group $\Gamma_{\langle\ModulusU\rangle}$ consists of the center $\{\pm\mathds{1}\}$ only. 
The center acts on chiral multiplets $\Phi$ and the superpotential as 
\begin{subequations}
\begin{align}
  \Phi&\xmapsto{~\gamma_0=-\mathds{1}~}(-1)^{\ModularWeight_\Phi}\,\rho(-\mathds{1})\,\Phi\eqdef\varepsilon_{\Phi}\,\Phi\;,\\
  \mathscr{W}&\xmapsto{~\gamma_0=-\mathds{1}~}(-1)^{\ModularWeight_{\mathscr{W}}}\,\mathscr{W}\;.
\end{align}  
\end{subequations}
This center symmetry holds independently of $\ModulusU$. 
We call $\Phi$ \emph{center-even} if $\varepsilon_{\Phi}=+1$ and \emph{center-odd} if $\varepsilon_{\Phi}=-1$. 
Note that the fact that $\varepsilon_{\Phi}$ can only take the values $\pm1$ follows from Schur's lemma and the fact that the symmetry is central. 

For odd $\ModularWeight_{\mathscr{W}}$ and an appropriate multiplier\footnote{If the superpotential transforms with a multiplier, $\mathscr{W}\mapsto\chi(\gamma)\,j^{\ModularWeight_{\mathscr{W}}}\,\mathscr{W}$, where $\chi$ is a character of \SL, the center acts on $\mathscr{W}$ by $\chi(-\mathds{1})\,(-1)^{\ModularWeight_{\mathscr{W}}}$. 
The center is then an $R$-symmetry if and only if $\chi(-\mathds{1})\,(-1)^{\ModularWeight_{\mathscr{W}}}=-1$. 
For odd $\ModularWeight_{\mathscr{W}}$, this holds for the trivial multiplier and, more generally, for all multipliers with $\chi(-\mathds{1})=1$.}, the center symmetry has order 4 since it acts on the superspace coordinates as 
\begin{equation}
 \theta\xmapsto{~\ZNR{4}~}\pm\ii\theta\;.  
\end{equation}
The sign ambiguity is the same as the one commented on in \Cref{ftn:no_order_1_or_2_R-symmetries}. 
In particular, every modular symmetry with $n_\mathrm{tori}=1$ and an appropriate multiplier leads to a $\ZNR{4}$ symmetry that is linearly realized, independently of $\ModulusU$. 
Note that the center symmetry, $\Generator{S}^2=-\mathds{1}$, has been dubbed modular spin in \cite{Liu:2025lym}.

\subsection{Modular Stabilizer Symmetries in \ac{4D} \acp{GUT}}

Modular symmetries provide us with a compelling way to address the flavor puzzle within the framework of modular flavor symmetries~\cite{Feruglio:2017spp}, see also the reviews~\cite{
Feruglio:2019ybq,
Almumin:2022rml,
Ding:2023htn,
Kobayashi:2023zzc}.
It is, therefore, natural to ask to which extent modular symmetries can be combined with \ac{4D} \acp{GUT}.

Given that the stabilizer symmetries of \Cref{sec:stabilizers} are linearly realized $R$-symmetries, our no-go theorem of \Cref{sec:No-Go_Theorem} applies.
This implies, in particular, that at the critical points in $\ModulusU$ moduli space, some exotic states become massless. 
Away from the fixed point, portions of the stabilizer symmetries are spontaneously broken, and the exotics can pick up masses. 
However, as stressed in \Cref{sec:center_and_Z_4_r}, the center symmetry remains unbroken.

\subsubsection{A No-Go Corollary for Modular Symmetries}

The abstract no-go theorem, \Cref{thm:abstract}, applies with $r=\gamma_{0}\in\Gamma_{\langle\ModulusU\rangle}$ and $\zeta^{2}=j_{0}^{\ModularWeight_{\mathscr{W}}}$, where $\gamma_{0}$ leaves the \ac{GUT}-breaking vacuum invariant. 
Here, $\Gamma_{\langle\ModulusU\rangle}$ denotes the stabilizer group defined above \Cref{eq:modular_trafos_stabilizer}.
Then, $\mathcal{F}_{-}$ consists of the fermions whose scalar partners are invariant under $\gamma_{0}$. 
Likewise, $\mathcal{F}_{+}$ consists of fermions whose scalar partners pick up the phase $j_{0}^{\ModularWeight_{\mathscr{W}}}$, together with the gauginos. 
Mass terms within $\mathcal{F}_{-}$ and within $\mathcal{F}_{+}$ are forbidden if and only if $N\nmid \ModularWeight_{\mathscr{W}}$, where $N$ denotes the order of $j_{0}$ (\lean{Modular.stabilizer\_phases}). 
In this case, the mass matrix has the block off-diagonal form~\eqref{eq:offdiag}. 
If $\gamma_{0}$ generates the full stabilizer, $N=2$, $4$ and $6$ at generic $\ModulusU$, at $\ModulusU_{0}=\ii$ and at $\ModulusU_{0}=\omegaup$, respectively.

\begin{corollary}\label{cor:stabilizer}
  The light spectrum of a \ac{4D} \SU{5} \ac{GUT} with unbroken \GSM{} cannot be the exact \ac{MSSM} with additional gauge singlets if an element of $\Gamma_{\langle\ModulusU\rangle}$ fixes the \ac{GUT}-breaking vacuum and multiplies the superpotential by a nontrivial phase. 
  More generally, this holds for any linearly and unitarily realized symmetry $g$, commuting with \SU{5}, with $\mathscr{W}\circ g=\eta\,\mathscr{W}$, $\eta\ne1$ and $g\,v=v$.
\end{corollary}

Note that \Cref{cor:stabilizer} covers $\ZNR{M}$, $\U{1}_{R}$, non-Abelian $R$-symmetries~\cite{Chen:2013dpa}, and the discrete $R$-symmetries of heterotic orbifolds~\cite{Kappl:2010yu,Nilles:2013lda,Liu:2025lym}.

\subsubsection{Modular \texorpdfstring{$\ZNR{4}$}{Z4R} Symmetry in \ac{4D} \acp{GUT}}
\label{sec:Z4R_exotics}

As we have seen in \Cref{sec:center_and_Z_4_r}, for odd superpotential weights there is a linearly realized $\ZNR{4}$ symmetry at a generic point in $\ModulusU$ moduli space. 
This leads to another corollary.  

\begin{corollary}[\lean{Modular.center\_bipartite}]\label{thm:center}
  Let $\ModularWeight_{\mathscr{W}}$ be odd, so that the superpotential is odd under the center with trivial multiplier. 
  Further, let the \ac{GUT}-breaking vacuum be center-even. 
  Then the conclusion of \Cref{cor:stabilizer} holds \emph{for every value of $\langle\ModulusU\rangle$}.
\end{corollary}

Hence, to avoid massless exotics, the \ac{GUT}-breaking vacuum has to break the modular $\ZNR{4}$, for instance through a center-odd \ac{GUT} Higgs. 
Indeed, for $\ModularWeight_{\mathscr{W}}=-1$ and a single center-odd $\rep{24}$-plet of weight $-3$, the $(\rep{8},\rep{1})$ and $(\rep{1},\rep{3})$ components acquire masses at generic $\ModulusU$.
This breaking occurs at $M_{\mathrm{GUT}}$ and may affect the predictions of modular flavor models.

Note that one cannot restore the modular $\ZNR{4}$ by combining it with an element of the center of \SU{5}.
This is because vacua of vanishing $5$-ality are invariant under the center of \SU{5}.
Hence, if the light spectrum is the exact \ac{MSSM}, no remnant of the modular $\ZNR{4}$ survives as an $R$-symmetry at low energies.
In particular, the modular $\ZNR{4}$ cannot be used to forbid the $\mu$ term.

\subsubsection{Parametrically Light Exotics near Critical Points}
\label{sec:light}

At the critical points $\ModulusU_{0}=\ii$ and $\ModulusU_{0}=\omegaup$ with $\ModularWeight_{\mathscr{W}}=-2$, the respective superpotential phases are $-1$ and $\omegaup^2$.  
Our no-go theorem applies at $\ModulusU_{0}$, provided that the full \ac{GUT}-breaking vacuum preserves the stabilizer. 
Near $\ModulusU_{0}$ it provides us with additional valuable insights. 
To see this, define~\cite{Feruglio:2021dte,Novichkov:2021evw} 
\begin{equation}\label{eq:expansion_parameter_u}
  u\defeq\begin{dcases}
    \dfrac{\ModulusU-\ii}{\ModulusU+\ii}&\text{for }\ModulusU_0=\ii\;,\\  
    \dfrac{\ModulusU-\omegaup}{\ModulusU-\omegaup^{2}}&\text{for }\ModulusU_0=\omegaup\;,  
  \end{dcases}
\end{equation}
which measures the departure of $\ModulusU$ from its critical value. 
Under the stabilizer transformations \eqref{eq:modular_trafos_stabilizer}, $u\mapsto j_0^{-2}\,u$, i.e.\ $u$ transforms like a field of modular weight $-2$. 
Hence $u$ carries $R$-charge, which we list in \Cref{tab:u_charges}. 

\begin{table}[htb]
  \centering
  \begin{tabular}{llll}
  \toprule
  $\ModulusU_{0}$ & $n_\mathrm{tori}=1$ & $n_\mathrm{tori}$ odd & $n_\mathrm{tori}$ even\\
  \midrule
  $\ii$ & $4$ under $\ZNR{8}$ & $4$ under $\ZNR{8}$ & \begin{tabular}[t]{@{}l@{}}$2$ under $\ZNR{4}$ for $n_\mathrm{tori}=2\bmod4$\\ --- for $4\mid n_\mathrm{tori}$\end{tabular}\\
  \midrule
  $\omegaup$ & $4$ under $\ZNR{12}$ & 
    \begin{tabular}[t]{@{}l@{}}
      $4$ under $\ZNR{12}$ for $n_\mathrm{tori}=1\bmod6$\\[1ex] 
      $8$ under $\ZNR{12}$ for $n_\mathrm{tori}=5\bmod6$\\[1ex] 
      $0$ under $\ZNR{4}$ for $3\mid n_\mathrm{tori}$
    \end{tabular} & \begin{tabular}[t]{@{}l@{}}
      $2$ under $\ZNR{6}$ for $n_\mathrm{tori}=2\bmod6$\\[1ex] 
      $4$ under $\ZNR{6}$ for $n_\mathrm{tori}=4\bmod6$\\[1ex] 
      --- for $6\mid n_\mathrm{tori}$\end{tabular}\\
  \bottomrule
  \end{tabular}
  \caption{$R$-charge of the expansion parameter $u$ under the linearly realized $\ZNR{M}$ symmetries of \Cref{tab:stabilizers}. 
  The $R$-charges are normalized such that $\theta$ has $R$-charge $1$ and $\mathscr{W}$ has $R$-charge $2$. 
  ``---'' indicates that there is no linearly realized $R$-symmetry.}
  \label{tab:u_charges}
\end{table}

By holomorphy, only nonnegative powers of $u$ appear in the superpotential.
Consider now a mass term whose residual charge differs from that of the superpotential by $\Delta r$. In order to be allowed, it must be multiplied by $u^{n}$, where $n$ is the least nonnegative integer with $q_u\,n= \Delta r\bmod M$ in $\Z{M}$.
Here, $q_u$ denotes the $R$-charge of $u$ under the linearly realized $\ZNR{M}$ as given in \Cref{tab:u_charges}.
Since the \ac{GUT}-breaking vacuum preserves the residual symmetry at $\ModulusU_{0}$, the \ac{GUT} Higgs $H$ has residual charge~$0$. 
For the mass term $\tr H^{2}$, the charge mismatch is therefore $\Delta r=2$.

As in the \ac{FN} picture, $u$ acts as a charged \ac{FN} spurion, so for a \ac{GUT} Higgs $H$ neutral under the residual symmetry, the leading allowed term $u^n\,\tr H^2$ is then determined by $\ZNR{M}$ charge conservation.
That is, $j_0^{-2n}=j_0^{\ModularWeight_{\mathscr{W}}}$, with $n$ the smallest nonnegative integer satisfying this condition. 
More specifically, we have
\begin{equation}
 m_{\mathrm{exotic}}\sim M_{\mathrm{GUT}}\,\lvert u\rvert^{n}\quad\text{with }
 n=\begin{dcases}
  1&\text{if } \ModulusU_{0}=\ii \text{ and } \ModularWeight_{\mathscr{W}}=2 \bmod 4\;,\\
  1&\text{if }\ModulusU_{0}=\omegaup \text{ and } \ModularWeight_{\mathscr{W}}=4\bmod 6\;,\\
  2&\text{if }\ModulusU_{0}=\omegaup \text{ and } \ModularWeight_{\mathscr{W}}= 2\bmod 6\;.
 \end{dcases}
\end{equation}
For odd $\ModularWeight_{\mathscr{W}}$ and $\ModulusU_{0}=\ii$, no $n$ exists. 
A center-even vacuum leaves the exotics massless for every $\ModulusU$ by \Cref{thm:center}. 
A center-odd vacuum, on the other hand, breaks the center. 
Every nontrivial subgroup of the stabilizer $\langle\Generator{S}\rangle\cong\Z{4}$ of $\ii$ contains the center. 
Hence a center-odd vacuum breaks the stabilizer completely. 
At $\ModulusU_{0}=\omegaup$, however, $\Generator{S}\,\Generator{T}$ generates a $\Z{3}$ subgroup that does not contain the center. 
At level one, a center-odd vacuum preserves this $\Z{3}$ if $3\mid\ModularWeight_{H}$. 
For $3\nmid\ModularWeight_{\mathscr{W}}$, the $\Z{3}$ multiplies the superpotential by $\omegaup^{2\ModularWeight_{\mathscr{W}}}\ne1$. 
Then, the leading allowed term $u^{n}\,\tr H^{2}$ has $n=1$ for $\ModularWeight_{\mathscr{W}}=1\bmod3$, and $n=2$ for $\ModularWeight_{\mathscr{W}}=2\bmod3$, so the exotics again have masses $m\sim M_{\mathrm{GUT}}\,\lvert u\rvert^{n}$. 
For even $\ModularWeight_{\mathscr{W}}$, this rule reproduces the values for $\ModulusU_{0}=\omegaup$ given above.

It is instructive to consider a simple toy model (\lean{Modular.toy\_model}). 
Take $\ModularWeight_{\mathscr{W}}=-2$ and a single $\rep{24}$ of weight $\ModularWeight_{H}=-4$. 
The superpotential is then given by
\begin{equation}
 \mathscr{W}_\mathrm{toy}=\mathcal{M}\,E_{6}(\ModulusU)\,\tr H^{2}+\lambda\,E_{4}(\ModulusU)\,E_{6}(\ModulusU)\,\tr H^{3}\;.
\end{equation}
We impose that the $F$-term of the \ac{GUT} Higgs vanishes, $F_{H}=0$.
For $\ModulusU\ne\ii$, the common factor $E_{6}(\ModulusU)$ cancels in the vacuum value $\langle H\rangle\propto \mathcal{M}/(\lambda\,E_{4})$, which stays finite as $\ModulusU\to\ii$. 
The masses of $(\rep{8},\rep{1})$ and $(\rep{1},\rep{3})$, however, are proportional to $E_{6}(\ModulusU)$, which has a simple zero at $\ii$. 
Numerically, for $\mathcal{M}=\lambda=1$, $\lvert m_{(\rep{8},\rep{1})}\rvert/\lvert\ModulusU-\ii\rvert\to66.6$. 
In particular, the masses of the exotics vanish exactly at $\ModulusU=\ii$.

What are the phenomenological consequences of these results? 
In modular flavor models, the proximity of the modulus to a fixed point is often used to generate fermion mass hierarchies~\cite{Feruglio:2021dte,Novichkov:2021evw,Chen:2025tby}. 
In fact, the expansion parameter $u$ of \Cref{eq:expansion_parameter_u} was originally used to parametrize such hierarchies. 
Our above discussion, therefore, implies that when such scenarios get combined with \ac{4D} \acp{GUT}, there will be parametrically light adjoint-type exotics. 
The suppression of their masses goes like $\lvert u\rvert^{n}$, just like the fermion masses. 
Generically, the presence of such states well below the \ac{GUT} scale is expected to impact gauge coupling unification, one of the very motivations of \acp{GUT}.

Let us also note that in string models winding modes become massless when $\ModulusU$ attains a critical value~\cite{Ibanez:1990ju,Dijkgraaf:1989hb,Beye:2014nxa,Li:2025bsr}.
However, in contrast to the above-mentioned exotics, these states do not carry \ac{SM} quantum numbers, and therefore do not affect gauge coupling unification.

\subsection{Modular Zeros beyond Symmetries}

There is a corner case in which no linearly realized $R$-symmetry exists, but the no-go theorem still applies.
It arises away from the fixed points, for even $\ModularWeight_{\mathscr{W}}$ and a multiplier with $\chi(-\mathds{1})=1$, e.g.\ the trivial one.
In this case, the center acts trivially on the superpotential, so it is not an $R$-symmetry.
Hence \Cref{cor:stabilizer} does not apply.
However, the proof of \Cref{thm:abstract} uses the $R$-symmetry only to obtain the block off-diagonal form~\eqref{eq:offdiag}.
As we shall see, so-called modular zeros~\cite{Liu:2025lym} can enforce this form even without a symmetry.

\begin{figure}[htb]
\centering
	\begin{tikzpicture}[box/.style={draw,semithick,rounded corners=2pt,align=center,text width=0.24\linewidth},
	desc/.style={align=left,text width=0.24\linewidth,font=\small},
	arr/.style={-{Stealth[length=4pt]}, black,thick,rounded corners=2pt}]
		\node[box](MZ){Modular zeros};
		\node[box,below=0.8cm of MZ](StZ){Structural zeros};
		\node[box,left=0.04\linewidth of StZ](TZ){Trivial zeros};
		\node[box,right=0.04\linewidth of StZ](SpZ){Sporadic zeros};
		\node[desc,below=1ex of TZ]{e.g.\ $\ModularWeight<0$\\[0.5ex] resemble holomorphic zeros};
		\node[desc,below=1ex of StZ]{e.g.\ modular spin\\[0.5ex] explain string selection rules};
		\node[desc,below=1ex of SpZ]{no obvious pattern\\[0.5ex] not due to linear symmetries};
		\draw[arr] (MZ) -| (TZ);
		\draw[arr] (MZ) -- (StZ);
		\draw[arr] (MZ) -| (SpZ);
	\end{tikzpicture}
\caption{Classification of modular zeros.}
\label{fig:modular_zeros}
\end{figure}
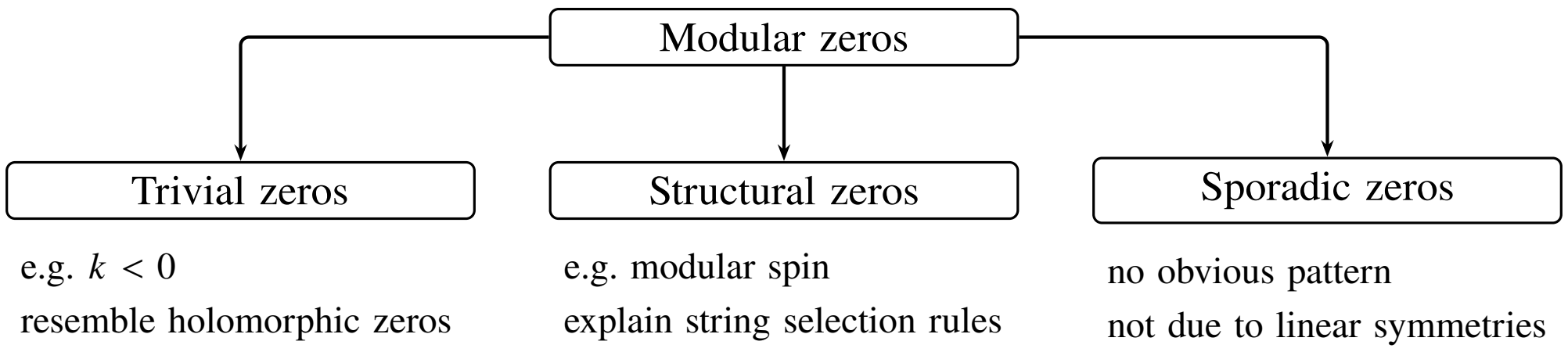

Modular zeros fall into three classes, see \Cref{fig:modular_zeros}.
Trivial zeros arise if the required weight has the wrong sign, e.g.\ $\ModularWeight<0$.
They are similar to the holomorphic zeros of \ac{FN}-type models~\cite{Leurer:1992wg,Leurer:1993gy}, which arise if the required charge has the wrong sign.
Structural zeros are due to modular spin, i.e.\ the center. 
These zeros can explain string selection rules that cannot be understood in terms of ordinary, linearly realized symmetries~\cite{Liu:2025lym,Knapp-Perez:2025cht}. 
Sporadic zeros follow no obvious pattern, and in general they, too, cannot be explained by linear symmetries.

In particular, modular zeros can forbid all mass terms among the fermions of the fields that attain \acp{VEV}, as well as all mass terms among their potential partners.
This is sufficient for~\eqref{eq:offdiag} to hold.

\begin{theorem}[\lean{Modular.gap\_bipartite}]\label{thm:gap}
	Consider level one with trivial multipliers and $\ModularWeight_{\mathscr{W}}=-2$.
	Suppose all chiral multiplets have modular weight $0$ or $-2$, and only fields of weight $0$ attain \acp{VEV}.
	Then, for every $\ModulusU$, the mass matrix has the block off-diagonal form~\eqref{eq:offdiag}.
	Here $\mathcal{F}_{-}$ consists of the fermions of weight $0$, and $\mathcal{F}_{+}$ of the fermions of weight $-2$ and the gauginos.
	Hence the conclusion of \Cref{thm:abstract} holds, although the center acts trivially on the superpotential.
\end{theorem}

Modular zeros can thus enforce~\eqref{eq:offdiag}.
In the example of \Cref{thm:gap}, an ordinary $\ZNR{4}$, under which the weight-$-2$ fields carry the same charge as $\mathscr{W}$, does the same.
However, the two scenarios differ in their implications for other couplings.
For a multiplet $P$ of weight $-2$, the $\ZNR{4}$ forbids $P^{4}$, whereas modular invariance allows $E_{6}(\ModulusU)\,P^{4}$.
More generally, no Abelian symmetry under which the vacuum is invariant allows $H\,P$, $P^{3}$ and $P^{4}$ but forbids $P^{2}$. 
This means that the modular zero scenario has quite unusual properties when compared to settings with ordinary symmetries.  

\subsection{Summary of the Modular Case}

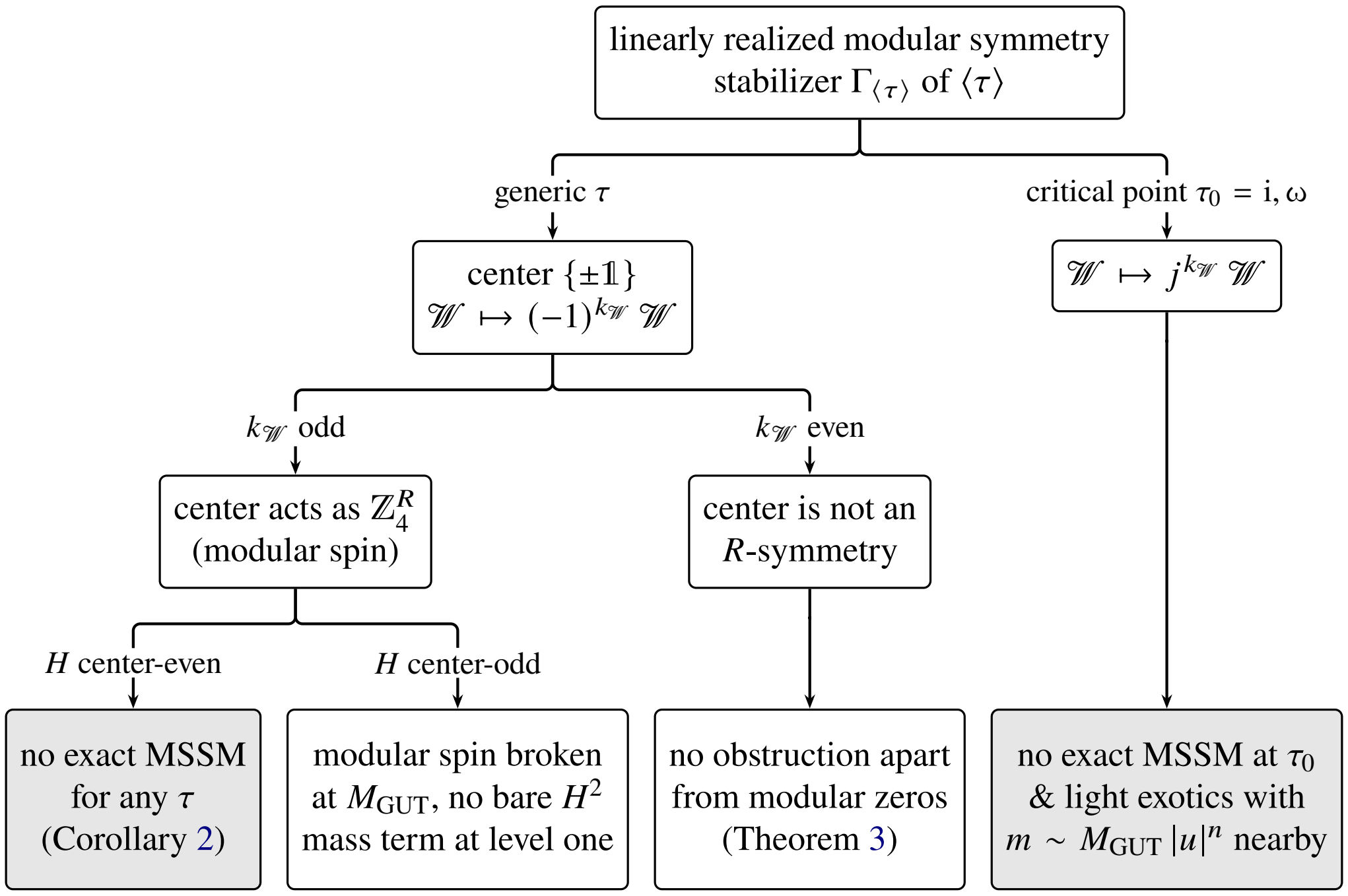
\begin{figure}[htb]
\centering
\begin{forest}
  forked edges,
  /tikz/el/.style={fill=white,inner sep=1.2pt,font=\small},
  for tree={
    draw=black,
	semithick,
    rounded corners=2pt,
    align=center,
    inner sep=4pt,
    edge={-{Stealth[length=4pt]}, black,thick,rounded corners=2pt},
    l sep=12mm,
    fork sep=3.5mm,
    s sep=2.5mm,
  },
  [{linearly realized modular symmetry\\ stabilizer $\Gamma_{\langle\ModulusU\rangle}$ of $\langle\ModulusU\rangle$}
    [{center $\{\pm\mathds{1}\}$\\ 
    $\mathscr{W}\mapsto(-1)^{\ModularWeight_{\mathscr{W}}}\,\mathscr{W}$}, 
    edge label={node[pos=0.73,el]{generic $\ModulusU$}}
      [{center acts as $\ZNR{4}$\\ (modular spin)}, edge label={node[pos=0.73,el]{$\ModularWeight_{\mathscr{W}}$ odd}}
        [{no exact \ac{MSSM}\\ for any $\ModulusU$\\ (\Cref{thm:center})}, 
        edge label={node[pos=0.73,el]{$H$ center-even}}, 
        fill=black!10, tier=leaf, anchor=north, minimum height=18mm]
        [{modular spin broken\\ at $M_{\mathrm{GUT}}$, no bare $H^{2}$\\ mass term at level one},
        edge label={node[pos=0.73,el]{$H$ center-odd}}, fill=white, tier=leaf, anchor=north, minimum height=18mm]
      ]
      [{center is not an\\ $R$-symmetry}, edge label={node[pos=0.73,el]{$\ModularWeight_{\mathscr{W}}$ even}}
        [{no obstruction apart\\ from modular zeros\\ (\Cref{thm:gap})}, fill=white, tier=leaf, anchor=north, minimum height=18mm]
      ]
    ]
    [{$\mathscr{W}\mapsto j^{\ModularWeight_{\mathscr{W}}}\,\mathscr{W}$},
    edge label={node[pos=0.73,el]{critical point $\ModulusU_{0}=\ii,\omegaup$}}
      [{
	  no exact \ac{MSSM} at $\ModulusU_{0}$\\ \& light exotics with\\ $m\sim M_{\mathrm{GUT}}\,\lvert u\rvert^{n}$ nearby}, fill=black!10, tier=leaf, anchor=north, minimum height=18mm]
    ]
  ]
\end{forest}
\caption{Consequences of the abstract no-go theorem, \Cref{thm:abstract}, for modular-invariant \ac{4D} \acp{GUT}. Shaded boxes mark the cases in which the exact \ac{MSSM} is excluded, and white boxes the cases in which no unbroken symmetry forbids it. $H$ denotes the \ac{GUT}-breaking Higgs and $u$ the local coordinate around the fixed point $\ModulusU_{0}$, see \Cref{sec:light}.}
\label{fig:cases}
\end{figure}

Let us summarize our discussion of modular symmetries in \ac{4D} \acp{GUT}. 
The different cases are illustrated in \Cref{fig:cases}.
\begin{enumerate}
 \item The no-go theorem applies to every linearly realized $R$-symmetry that leaves the vacuum invariant and multiplies the superpotential by a nontrivial phase (\Cref{cor:stabilizer}). 
 For modular symmetries, the stabilizer of $\langle\ModulusU\rangle$ falls into this class. 
 At generic $\ModulusU$, only the center remains. 
 Therefore, the symmetry-based no-go theorem holds for all $\ModulusU$ if and only if $\mathscr{W}$ is center-odd and the vacuum center-even (\Cref{thm:center}).
 \item Near a fixed point the no-go theorem turns into a prediction of light exotics. 
 Their masses are suppressed by the powers $\lvert u\rvert^{n}$, which are determined by the residual charges (\Cref{sec:light}).
 \item In the absence of linearly realized symmetries, there can still be modular zeros which prevent us from giving exotics masses (\Cref{thm:gap}). 
\end{enumerate}

\section{Concluding Remarks}\label{sec:conclusions}

The combination of Lean and \ac{AI} has revolutionized mathematics. 
In this work, we used Lean to formalize no-go theorems in high-energy physics. 
This has allowed us to double-check the validity of these theorems and extend their scope. 
In the process, we identified and closed a gap in an earlier proof.

We also extended the analysis to modular symmetries, which we have argued to be inevitably $R$-symmetries in any supersymmetric framework that also describes gravity. 
The no-go theorems then led to interesting corollaries for modular symmetries. 
In the context of \ac{4D} \acp{GUT}, modular symmetries imply the existence of exotic states that become massless at the critical points. 
This is because modular $R$-symmetries are linearly realized at these points.
In the framework of modular flavor symmetries, the modulus is often required to settle close to some critical point in order to accommodate fermion mass hierarchies. 
The no-go theorems then imply the existence of exotic states that become parametrically light. 
These states appear in incomplete \ac{GUT} multiplets, and thus have the potential of ruining gauge coupling unification, one of the main motivations for \acp{GUT}. 

We have also seen that higher-dimensional models of grand unification evade the no-go theorems. 
This means that an \ac{MSSM} model with $R$- and modular symmetries is compatible with unification provided that the \ac{GUT} symmetry gets broken by compactification. 
In particular, string theory can give rise to such scenarios.

\subsection*{Acknowledgments}
X.-G.~L.\ was supported by the Universidad Nacional Autónoma de México Postdoctoral Program (POSDOC). 
A.S.\ was supported by the National Science Foundation Graduate Research Fellowship Program. 
This material is based upon work supported by the National Science Foundation Graduate Research Fellowship Program under Grant No.\ DGE-2235784. 
We acknowledge help by Anthropic's Claude to generate Lean code, and to fix typographical and grammatical errors.

\appendix

\section{What Has Been Formalized}\label{sec:lean}

We have formalized the abstract no-go theorem, \Cref{thm:abstract}, and its application to \SU5 in the proof assistant Lean~4, using the mathematics library Mathlib.
Lean accepts a proof only if every step can be checked mechanically against the definitions.
The formalization therefore tests whether the argument of \Cref{sec:No-Go_Theorem} really follows from the stated hypotheses. 
It also makes these hypotheses explicit.
The no-go theorem is formalized in the file \texttt{RNoGoAbstract.lean}, which is provided as supplementary material and described in \Cref{app:Lean}.
The modular selection rules of \Cref{sec:modular} are formalized in a separate file, \texttt{Modular.lean}
The command \texttt{\#print axioms} lists every axiom a theorem ultimately relies on.
For all main results, it reports only the three standard axioms of Lean's logic, \texttt{propext}, \texttt{Classical.choice} and \texttt{Quot.sound}.
\Cref{tab:lean} lists the main statements along with the corresponding Lean theorems.

\subsection{What Is Proved}

Let us go through the argument of \Cref{sec:No-Go_Theorem}.
The selection rule is proved.
If $\zeta^{2}\neq1$, then also $\zeta^{-1}\zeta^{-1}\neq1$ and $\zeta\,\zeta\neq1$.
Hence an $r$-invariant mass term can only pair $\mathcal{F}_-$ with $\mathcal{F}_+$.
The counting identity~\eqref{eq:ind_F} is proved from rank-nullity, separately for every weight block $\mathfrak{M}_\chi$.
The Weyl invariance~\eqref{eq:ind_weights} is proved as in the text.
A representative $n$ of $w\in\Weyl(B,T_0)$ maps the weight space $V(\chi)$ onto $V(w\cdot\chi)$.
Hence the two weight spaces have the same dimension.
\Cref{thm:abstract} then follows, because a Weyl-invariant function cannot take the value $-1$ at $\beta$ and $0$ at $w\beta$.
Finally, the \SU5 contradiction~\eqref{eq:contradiction} is proved as a special case.
Here the weights are elements of $\mathds{Z}^{5}$, $S_5$ permutes them, and $\Delta_0$ is the set of the eight roots of \GSM.
$5$-ality shows that no \ac{MSSM} field sits at a root.
A singlet weight is a multiple of $(1,1,1,1,1)$ and hence never a root.
What remains are three statements about the massless spectrum.
First, the massless fermions in $\mathcal{F}_-$ are \ac{MSSM} fields or singlets.
Second, there is exactly one massless fermion in $\mathcal{F}_+$ at the weight $-(e_0-e_1)$, namely a gluino.
Third, there is no massless fermion in $\mathcal{F}_+$ at the weight $-(e_0-e_3)$.
The Lean theorem shows that these three statements are contradictory.

\subsection{Modular Selection Rules}

The file \texttt{Modular.lean} formalizes the selection rules of \Cref{sec:modular} at level one with trivial multiplier.
A coupling is defined to be allowed if the space of modular forms of the required weight is nonzero.
Which of these spaces vanish is taken from Mathlib.
On this basis, the file proves the selection rules behind \Cref{cor:stabilizer,thm:center} and \Cref{thm:gap}, as well as those of the toy model of \Cref{sec:light}.
These results establish the block off-diagonal form~\eqref{eq:offdiag}.
They have not yet been combined with the Lean version of \Cref{thm:abstract} into a single theorem.

\subsection{What Is Assumed}

Mathlib does not yet contain the structure theory of compact Lie groups, such as maximal tori and the Weyl groups $N_B(T_0)/C_B(T_0)$.
The corresponding input therefore enters the Lean theorems as explicit hypotheses, never as axioms.
These hypotheses are the fields of the structure \lean{RFermionSystem} and the hypotheses of the theorem \lean{no\_go}.
There are three kinds of input.

First, $\mathcal{F}_-$ and $\mathcal{F}_+$ are finite-dimensional representations of $T_0$.
Further, for every $w\in\Weyl(B,T_0)$ there are invertible maps on $\mathcal{F}_-$ and on $\mathcal{F}_+$ that intertwine the action of $t\in T_0$ with that of $n\,t\,n^{-1}$, where $n$ represents $w$.
It is then proved that these maps send the weight space of $\chi$ onto that of $w\cdot\chi$.
This replaces \Cref{ass:finitely_many_multiplets,ass:r_comm}.
Indeed, by \Cref{ass:finitely_many_multiplets} $\mathcal{F}$ is a finite-dimensional $G$-representation.
Since $r$ commutes with $B$, $\mathcal{F}_-$ and $\mathcal{F}_+$ are separately $B$-representations, and representatives of $w$ in $B$ provide the required maps.

Second, the mass operator is given directly as the family of maps $\mathfrak{M}_\chi\colon\mathcal{F}_-(\chi)\to\mathcal{F}_+(-\chi)^\vee$.
That is, the block form~\eqref{eq:offdiag} and the $T_0$-invariance of \Cref{ass:vac} are built into the setup.
They are not derived from a supersymmetric Lagrangean.

Third, the Weyl group acts on the weights by additive maps, as $(w\cdot\chi)(t)=\chi(n^{-1}\,t\,n)$.
In Lean this is the hypothesis \lean{CharCompat} of \lean{no\_go}.
It is the only fact from Lie theory that the proof of \Cref{thm:abstract} uses.
No further property of the roots is needed.
Indeed, $\Delta_0$ enters only through $\beta\in\Delta_0$ and $w\cdot\beta\notin\Delta_0$.
For \SU5 the weights, the action of $S_5$ and the torus are defined explicitly, and \lean{CharCompat} is proved.
Hence the \SU5 result rests only on the first two kinds of input and on the three statements about the massless spectrum.

The conditions (a) and (b) of \Cref{thm:abstract} are hypotheses of the Lean theorem.
In (b), the gaugino count is written as $1$ on $\Delta_0$ and $0$ elsewhere, which is equivalent because the root spaces of $G_0$ are one-dimensional.

\subsection{What Is Not Formalized}

The expression~\eqref{eq:weylBT0} for $\Weyl(B,T_0)$ and the connectedness of $C_B(T_0)$ are not formalized, and neither is the derivation of the fermion mass matrix at a supersymmetric vacuum.
For a general group $B$ the existence of a pair $(\beta,w)$ is not formalized. 
For \SU5 the pair $\beta=e_0-e_1$, $w=(1\,3)$ is checked explicitly.
The application to larger gauge groups in \Cref{sec:Larger} is proved on paper only, as is the discussion of non-standard embeddings.
In the modular case, the zero $E_{6}(\ii)=0$ and the scaling of the masses of the light exotics in \Cref{sec:light} are not formalized.

\subsection{Lessons}

Two lessons from the formalization deserve mention.
First, the order of the $R$-symmetry is used in a single place, through $\zeta^{2}\neq1$ in the selection rule.
The argument is therefore the same for every \ZNR{M} with $M\notin\{1,2\}$. 
It also applies to $\U{1}_R$.
Second, the vacuum enters in only three ways.
It enters through its symmetries, through the requirement that the light chiral spectrum be the \ac{MSSM} plus singlets, and through the gaugino count~(b).
The gaugino count means one massless gluino at the weight $-(e_0-e_1)$ of an unbroken root, and no massless fermion at the weight $-(e_0-e_3)$ of its broken Weyl image.
Neither $F$- and $D$-flatness nor the counting of Goldstone modes is needed.

\begin{table}[tbp]
	\centering
	\begin{tabular}{@{}p{0.34\textwidth}p{0.44\textwidth}p{0.16\textwidth}@{}}
		\toprule
		Statement & Lean theorem & Status \\
		\midrule
		Selection rule behind~\eqref{eq:offdiag} & \lean{selection\_rule} & proved \\
		Counting identity~\eqref{eq:ind_F} & \lean{RFermionSystem.index\_eq} & proved \\
		Weyl invariance~\eqref{eq:ind_weights} & \lean{finrank\_weightSpace\_act}, \lean{RFermionSystem.index\_act} & proved \\
		\Cref{thm:abstract} & \lean{no\_go}, \lean{no\_go\_of\_Lplus\_ne} & proved \\
		\SU5 contradiction~\eqref{eq:contradiction} & \lean{SU5.no\_exact\_MSSM} & proved \\
		Selection rule of \Cref{cor:stabilizer} & \lean{Modular.stabilizer\_phases} & proved \\
		Selection rule of \Cref{thm:center} & \lean{Modular.center\_bipartite} & proved \\
		Selection rule of \Cref{thm:gap} & \lean{Modular.gap\_bipartite} & proved \\
		Selection rules of the toy model & \lean{Modular.toy\_model} & proved \\
		\Cref{eq:weylBT0}, \Cref{sec:Larger}, light exotics (\Cref{sec:light}) & --- & not formalized \\
		\bottomrule
	\end{tabular}
	
	\medskip
	\caption{Main statements of the paper and the corresponding Lean theorems in \texttt{RNoGoAbstract.lean} and, for the theorems with the prefix \lean{Modular}, in \texttt{Modular.lean}.}
	\label{tab:lean}

\end{table}

\section{Lean Formalization of the Abstract No-Go Theorem}\label{app:Lean}

In this appendix we show the central parts of the Lean file, so that the reader can see how the argument of \Cref{sec:No-Go_Theorem} reads in Lean.
The complete file is provided as supplementary material, \texttt{RNoGoAbstract.lean}.
It was checked with Lean~4.33.0 and Mathlib commit \texttt{db584cd6d46c92f209a44c0f1c829460d327499d}.
Each box below reproduces a range of lines of this file without any change, and the title of the box gives the range. 

\paragraph{Reading the Lean Code.}
A few words on the syntax are in order.
A \lean{theorem} lists its hypotheses as named arguments in parentheses, then a colon and the conclusion, and then the proof after \lean{:=}.
Most of our theorems have the conclusion \lean{False}.
That is, they state that their hypotheses cannot all hold, which is how a no-go theorem looks in Lean.
A \lean{structure} collects data together with the conditions they satisfy, and an argument such as \lean{(S : RFermionSystem act chr wt)} reads ``let $S$ be a fermion system.''
Text between \lean{/-} and \lean{-/} is documentation and is not checked by Lean.

We follow the notation of \Cref{sec:No-Go_Theorem} as far as possible. 
\lean{X} is the set of $T_0$-weights, \lean{W} is the group $\Weyl(B,T_0)$, and \lean{act}~$w$~$\chi$ stands for $w\cdot\chi$.
The fermion spaces $\mathcal{F}_-$ and $\mathcal{F}_+$ are called \lean{Fm} and \lean{Fp}.
In the comments of the file they are labeled by the fermion $R$-charges $f=-1$ and $f=+1$ that they have for a \ZNR{M} symmetry.
The letter \lean{K} denotes the field of numbers, for instance $\mathds{C}$, and not a group.
One convention needs care.
\lean{Lminus}~$\chi$ is $L_-(\chi)$, the number of massless fermions in $\mathcal{F}_-$ at weight $\chi$. \lean{Lplus}~$\chi$, on the other hand, is the dimension of the cokernel of $\mathfrak{M}_\chi$, i.e.\ $L_+(-\chi)$, the number of massless fermions in $\mathcal{F}_+$ at weight $-\chi$.

\paragraph{The Fermion System.}
Recall that the argument of \Cref{sec:No-Go_Theorem} uses three ingredients, the two fermion sectors $\mathcal{F}_\pm$ with the action of $T_0$, the action of the Weyl group, and the mass blocks $\mathfrak{M}_\chi$.
The structure \lean{RFermionSystem} collects exactly these.
Its fields are the finite-dimensional spaces \lean{Fm} and \lean{Fp}, cf.\ \Cref{ass:finitely_many_multiplets}, the torus actions \lean{$\rho$m} and \lean{$\rho$p}, the invertible maps \lean{nm}~$w$ and \lean{np}~$w$ by which a representative of $w$ acts, together with the condition that they intertwine the torus action, and the mass blocks \lean{M}.
The maps \lean{nm} and \lean{np} encode that the fermions come in $B$-multiplets and that $r$ commutes with $B$, \Cref{ass:r_comm}.
Note that \lean{M}~$\chi$ is a map from $\mathcal{F}_-(\chi)$ to the dual of $\mathcal{F}_+(-\chi)$ by construction.
That is, the block form~\eqref{eq:offdiag} and the $T_0$-invariance of \Cref{ass:vac} are built into the definition.
The algebraic step behind~\eqref{eq:offdiag}, that $\zeta^2\neq1$ implies $\zeta^{-2}\neq1$, is proved separately as \lean{selection\_rule}.

\leanexcerpt{RNoGoAbstract.lean}{139}{167}{RFermionSystem}

\paragraph{The Index and Its Weyl Invariance.}
Next, the file defines \lean{Lminus}, \lean{Lplus} and the index $\ind\mathfrak{M}_\chi=L_-(\chi)-L_+(-\chi)$.
The theorem \lean{index\_eq} is the counting identity~\eqref{eq:ind_F}.
Its proof applies rank-nullity to $\mathfrak{M}_\chi$, uses that a space and its dual have the same dimension, and finishes with \lean{omega}, a tactic that solves linear equations between integers.
The theorem \lean{index\_act} is the Weyl invariance~\eqref{eq:ind_weights}.
It rewrites both sides with \lean{index\_eq} and then uses that weight multiplicities are Weyl invariant, \lean{finrank\_weightSpace\_act} (lines 115--121).
The latter is proved exactly as in \Cref{sec:No-Go_Theorem}.
A representative $n$ of $w$ maps $V(\chi)$ onto $V(w\cdot\chi)$.
The hypothesis \lean{hc : CharCompat act chr wt} is the formula $(w\cdot\chi)(t)=\chi(n^{-1}\,t\,n)$ that relates the action of $w$ on weights to its action on the torus.

\leanexcerpt{RNoGoAbstract.lean}{179}{204}{Lminus, Lplus, index, index\_eq, index\_act}

\paragraph{The No-Go Theorem.}
The proof of \Cref{thm:abstract} has two steps.
The first is the elementary statement \lean{orbit\_contradiction} (lines 23--27).
It states that a function that is invariant under a group action cannot take different values at $x$ and $w\cdot x$.
The theorem \lean{no\_go\_of\_Lplus\_ne} applies it to the index.
If $L_-$ vanishes at $\beta$ and at $w\cdot\beta$, the index equals $-$\lean{Lplus} at both weights, so \lean{Lplus} must take the same value there.

\leanexcerpt{RNoGoAbstract.lean}{215}{225}{no\_go\_of\_Lplus\_ne}

The second step is the theorem \lean{no\_go}, which is \Cref{thm:abstract} itself.
Read from top to bottom, it says the following.
Let $S$ be a fermion system whose Weyl action satisfies \lean{CharCompat}, let $\Delta_0$ be a set of weights, let $\beta\in\Delta_0$, and let $w$ be an element of the Weyl group with $w\cdot\beta\notin\Delta_0$.
Assume condition (a), i.e.\ \lean{Lminus} vanishes at $\beta$ and at $w\cdot\beta$, and condition (b) in the form \lean{Lplus}~$\chi=1$ for $\chi\in\Delta_0$ and $0$ otherwise, for $\chi\in\{\beta,w\cdot\beta\}$. 
Then we arrive at a contradiction.
Since $\Delta_0=-\Delta_0$ and root spaces are one-dimensional, this form of (b) is condition (b) of \Cref{thm:abstract}.
The input from Lie theory enters only through \lean{CharCompat} and the intertwiners of the fermion system, and no property of $\Delta_0$ is needed beyond $\beta\in\Delta_0$ and $w\cdot\beta\notin\Delta_0$.
The proof evaluates (b) at the two weights, which gives \lean{Lplus}~$\beta=1$ and \lean{Lplus}~$(w\cdot\beta)=0$, and applies \lean{no\_go\_of\_Lplus\_ne}.
The line \lean{open Classical in} just above the box (line 227) allows the case distinction in (b) for an arbitrary set $\Delta_0$.

\leanexcerpt{RNoGoAbstract.lean}{228}{239}{no\_go}

\paragraph{The \SU5 Case.}
Finally, let us see how the \SU5 result follows.
The weights are the elements of $\mathds{Z}^5$, i.e.\ the characters of the diagonal torus of \U5.
Weights that differ by a multiple of $(1,1,1,1,1)$ agree on the torus of \SU5, and these multiples are the singlet weights.
\lean{W} is $S_5$ acting by permutations, and $\Delta_0$ is the set of the eight roots of \GSM.
The theorem \lean{SU5.no\_exact\_MSSM} has three hypotheses. 
\lean{minus\_content} says that every massless fermion in $\mathcal{F}_-$ has the weight of an \ac{MSSM} field or of a singlet.
\lean{plus\_unbroken} says that there is exactly one massless fermion in $\mathcal{F}_+$, the gluino, at weight $-(e_0-e_1)$.
\lean{plus\_broken} says that there is none at weight $-(e_0-e_3)$.
The proof first shows that $L_-$ vanishes at both roots.
No \ac{MSSM} weight is a root because of $5$-ality, and no singlet weight is a root because a multiple of $(1,1,1,1,1)$ never is.
It then applies \lean{no\_go} with $\beta=e_0-e_1$ and $w=(1\,3)$.
This is the contradiction~\eqref{eq:contradiction}.
Here the action of $S_5$ on the weights and on the torus is defined explicitly and \lean{CharCompat} is proved as \lean{SU5.su5\_charCompat}, so no hypothesis about Lie theory remains.

\leanexcerpt{RNoGoAbstract.lean}{477}{497}{SU5.no\_exact\_MSSM}

\bibliographystyle{utphys}
\bibliography{NoGo_R_and_modular}

\end{document}